\documentclass{emulateapj}    

\def\ajbib{

}

\def\figaa{
\begin{figure*}
\plotone{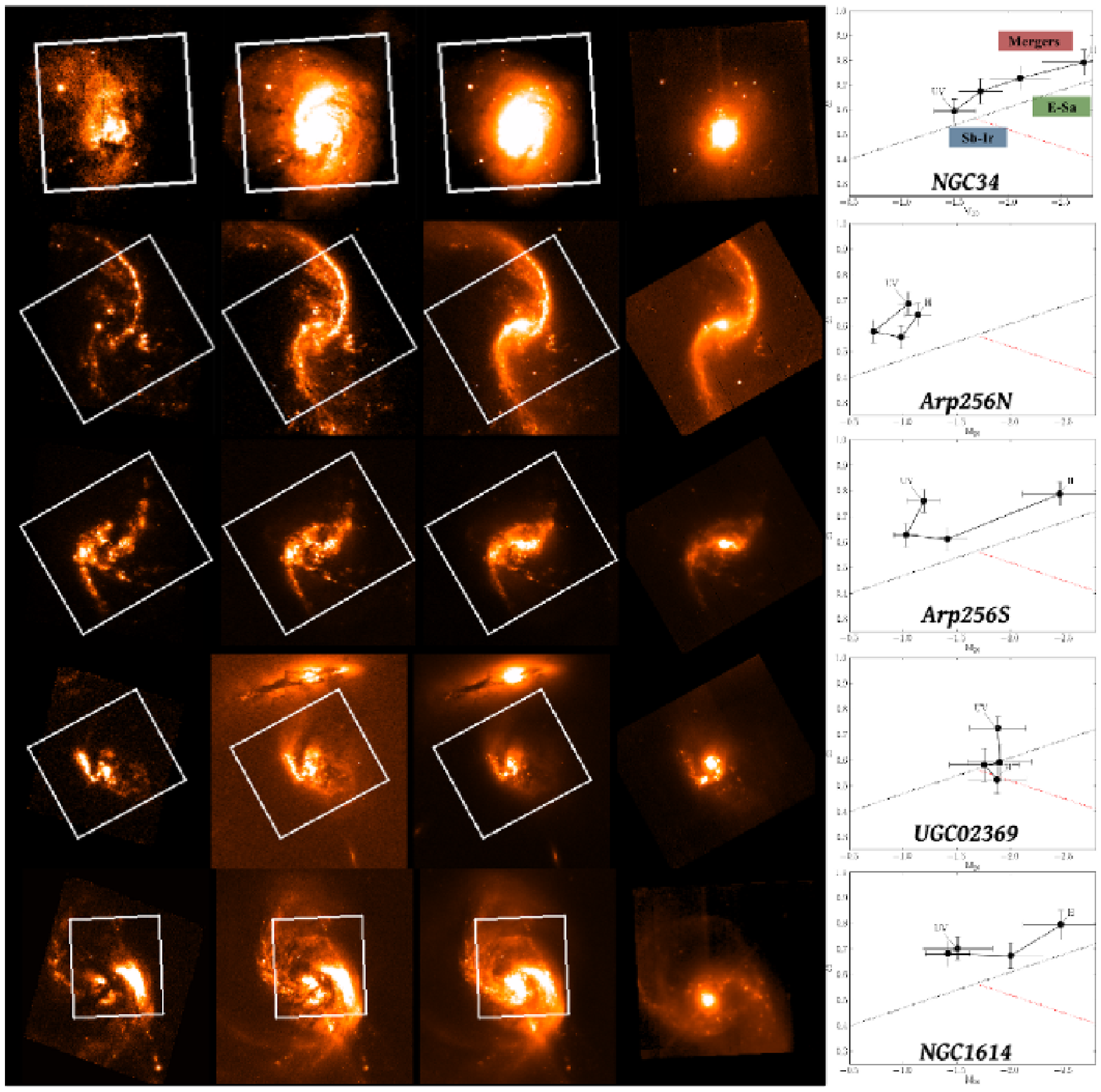}
\caption{The GOALS subsample in FUV-, B-, I-, and H-bands (left to right). The H-band footprint is outlined in a white box region. Individual plots of the G-\M20~ morphologies are shown to trace the behavior with wavelength. The individual G-\M20~ plots to the right of the image cutouts are the FUV-, B-, I-, H-band values (UV and H are annotated).}\label{fig1}
\end{figure*}
}

\def\figab{
\begin{figure*}
\plotone{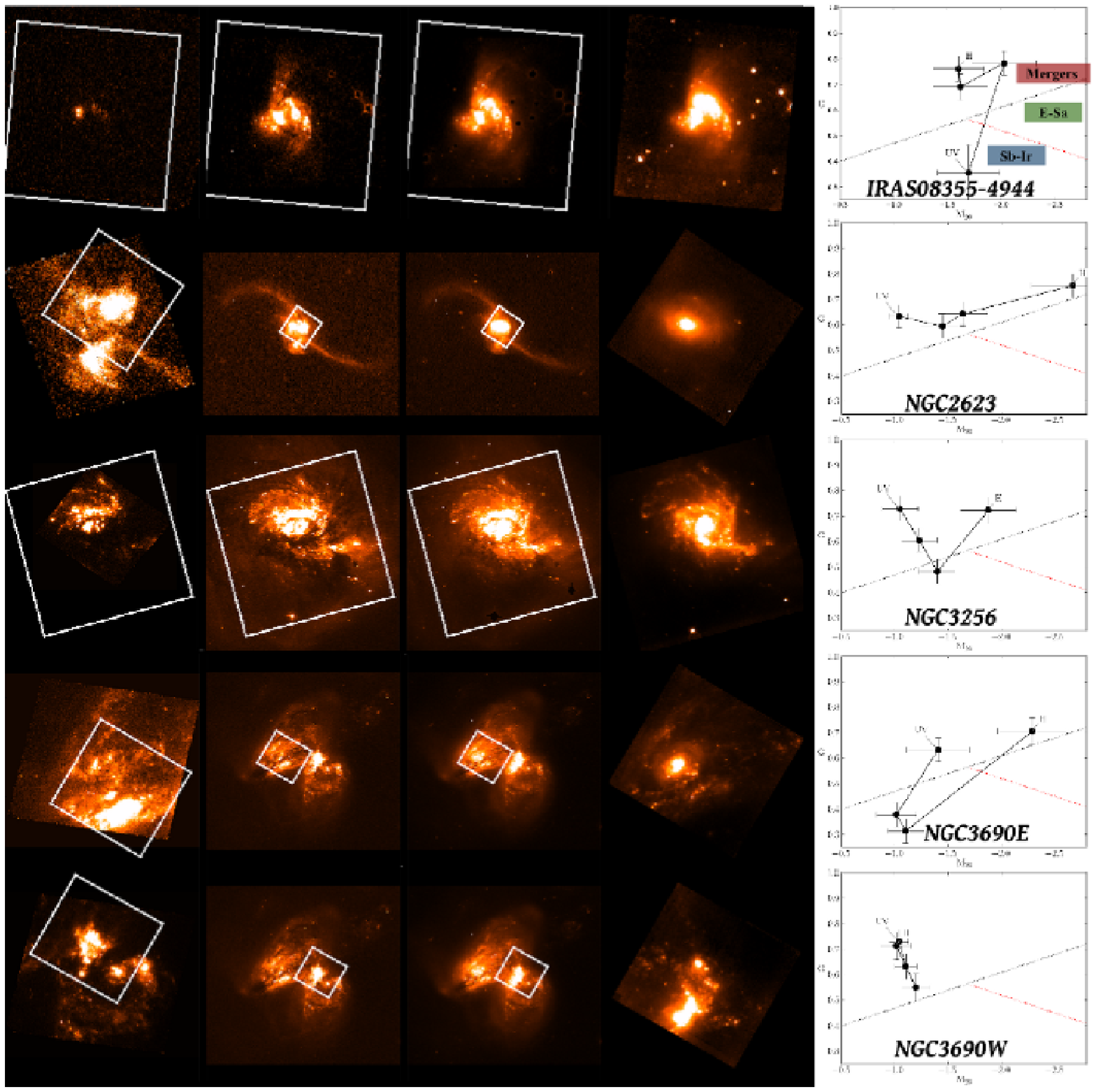}
\caption{Continued from Figure \ref{fig1}}\label{fig1a}
\end{figure*}
}

\def\figac{
\begin{figure*}
\plotone{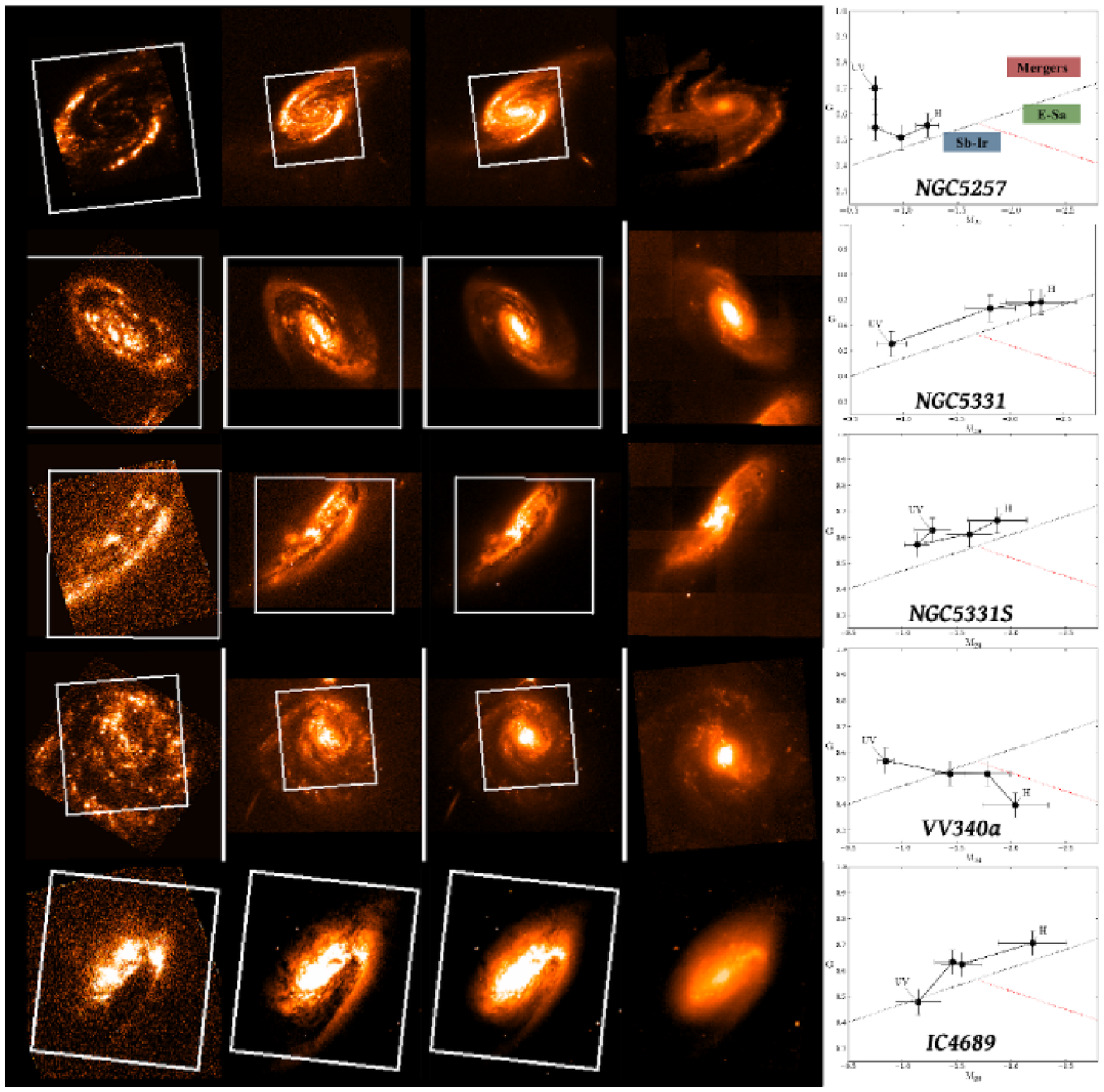}
\caption{Continued from Figure \ref{fig1}}\label{fig1b}
\end{figure*}
}
\def\figad{
\begin{figure*}
\plotone{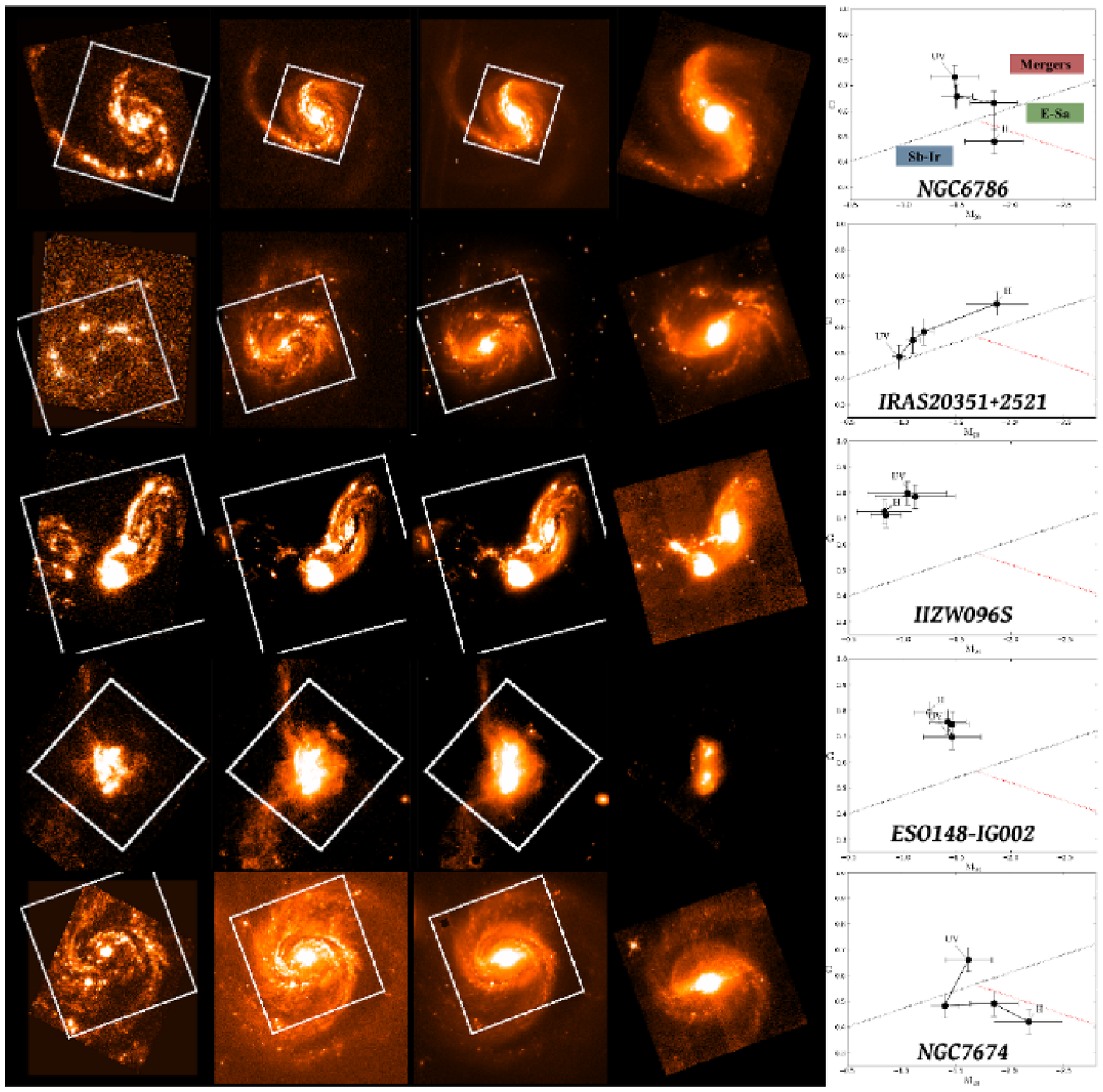}
\caption{Continued from Figure \ref{fig1}}\label{fig1c}
\end{figure*}
}

\def\figc{
\begin{figure}

\plotone{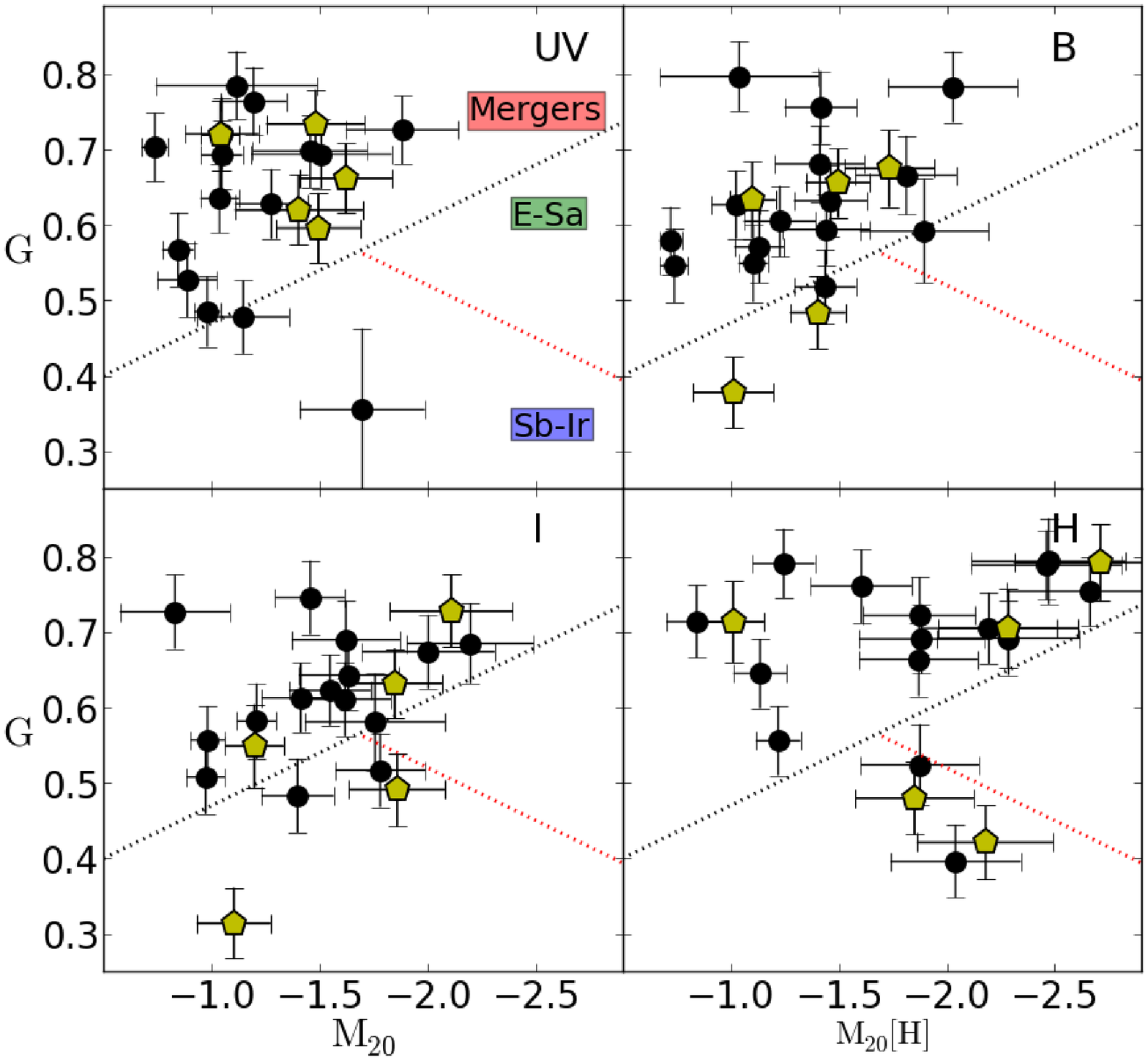}
\caption{G-\M20~ at multiple wavelengths (UV, B, I, and H, as labeled) show that, within the \citet{lotz08b} lines, the quantitative morphology is consistent from UV to near-IR. Most of the objects are in the merger-designated region and move around in that relative space. Figures \ref{fig1}-\ref{fig1c} show this for each object, where only a few cross the merger line. The yellow pentagons are the galaxies known to be hosting AGN (NGC 34, NGC 3690E, NGC3690W, NGC 6786, NGC 7674). There is no clear correlation between an AGN, dominating the morphology as it becomes unveiled at longer wavelengths.}\label{fig3}
\end{figure}
}

 \def\figca{
 \begin{figure}

 \plotone{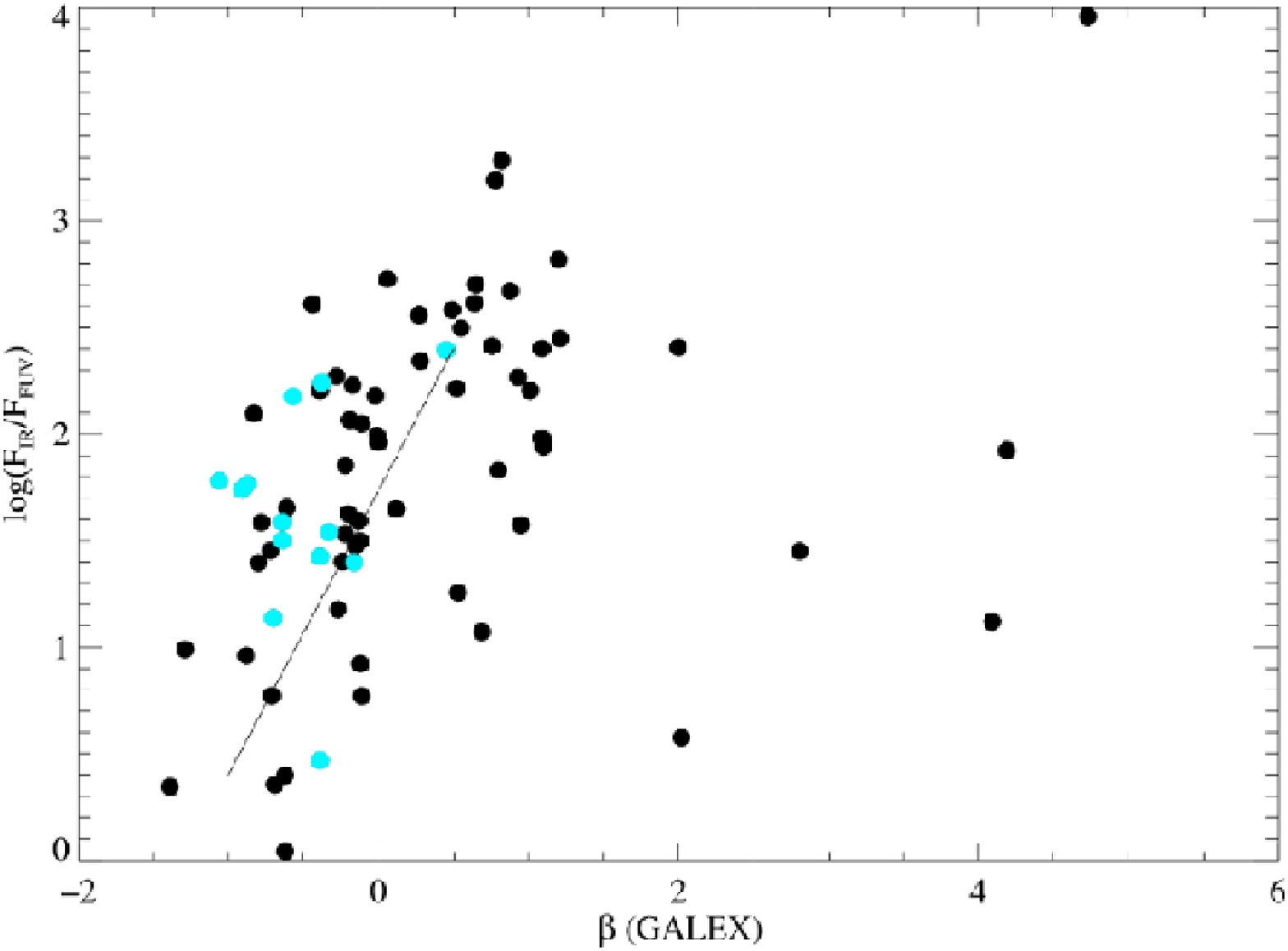}
 \caption{IRX-$\beta$ for GOALS LIRGs. The black filled circles are from \cite{how10}, and the cyan points are the LIRGs in the current paper. The black line is the starburst relation adapted from the fit to a \cite{meu99} $10^8$ year-old starburst population. While the cyan points span a large range in IRX, including a number that fall significantly above the starburst correlation, they do not cover the full range in beta seen in the parent GOALS sample.
 }\label{fig3a}
 \end{figure}
 }

 \def\figcb{
 \begin{figure}

 \plotone{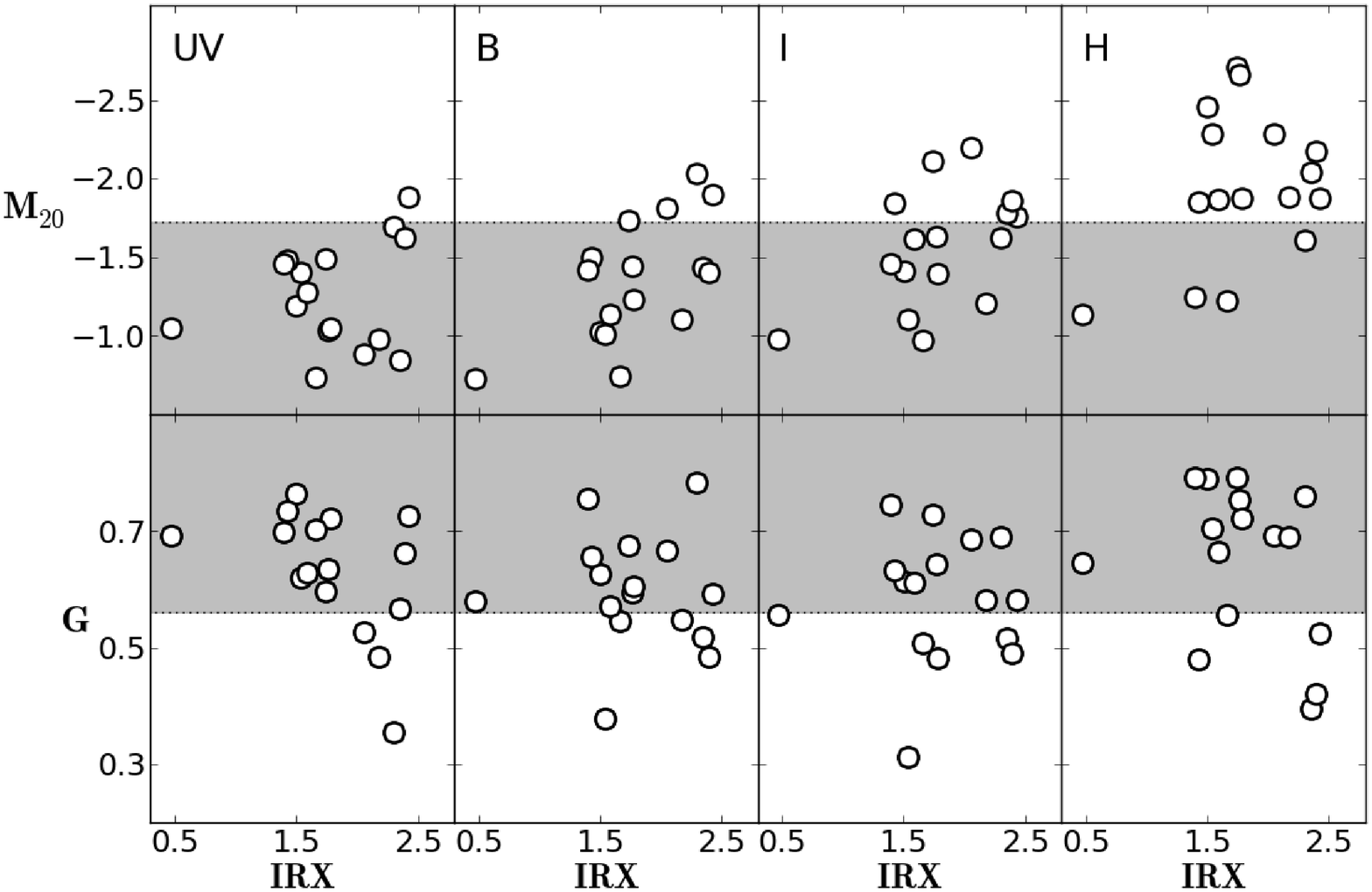}
 \caption{
 IRX-\M20 and G for 16 objects with available IRX in the UV-, B-, I- and H-bands. The grey-shaded regions roughly define more merger- or clumpy disk-like composition based on the G-\M20 boundary.
 }\label{fig4}
 \end{figure}
 }

\def\fige{
\begin{figure}

\plotone{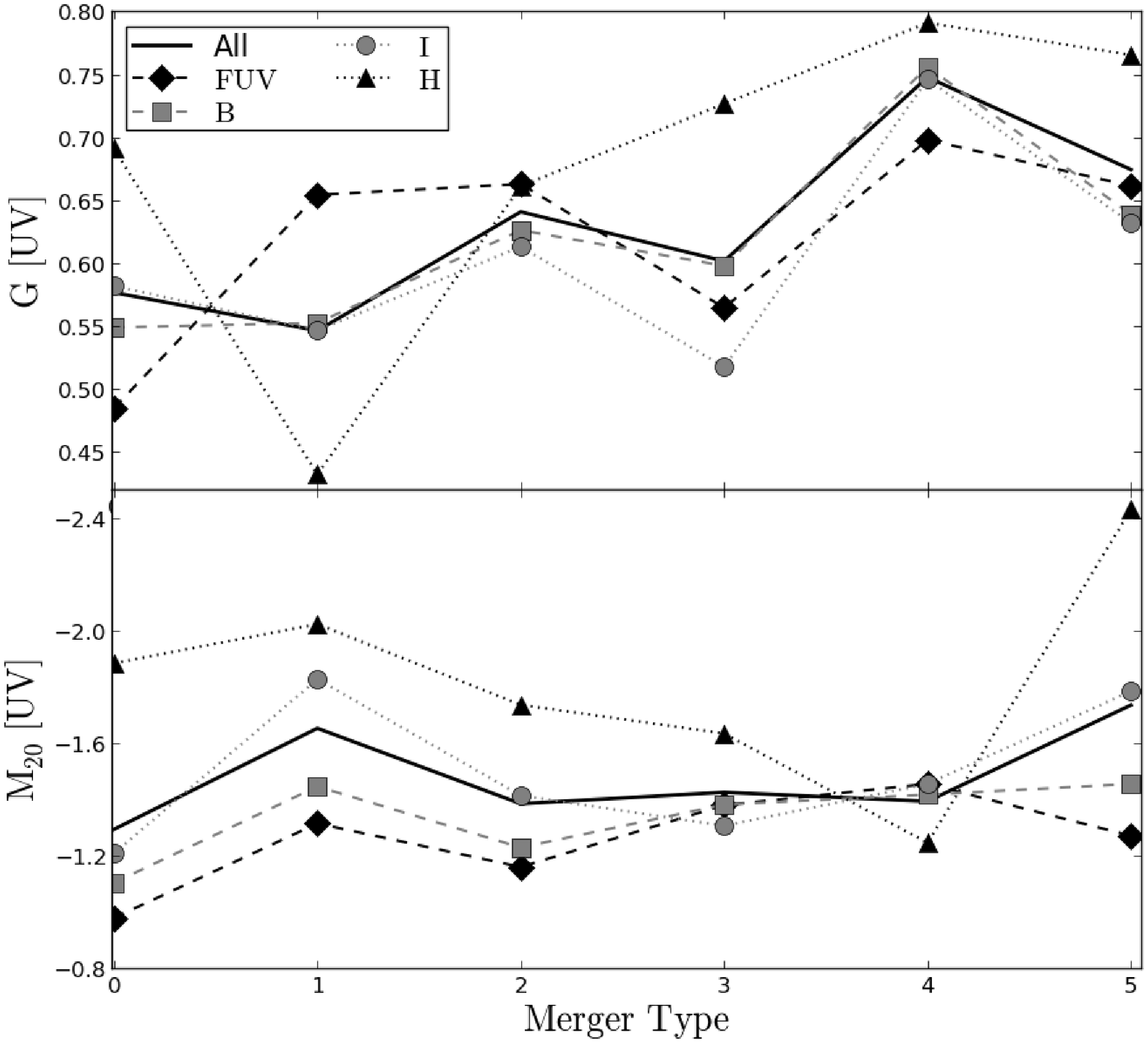}
\caption{
Plots are of the average G (top) and \M20 (bottom) values in different wavebands compared to the visual classification type (Table \ref{tableA}). Black diamonds, grey squares, grey circles and black triangles are the mean values in the FUV-, B-, I- and H-bands, respectively. The black solid line is the average for all values in that classification type.
}\label{fig5}
\end{figure}

}
\def\figg{
\begin{figure*}
\plotone{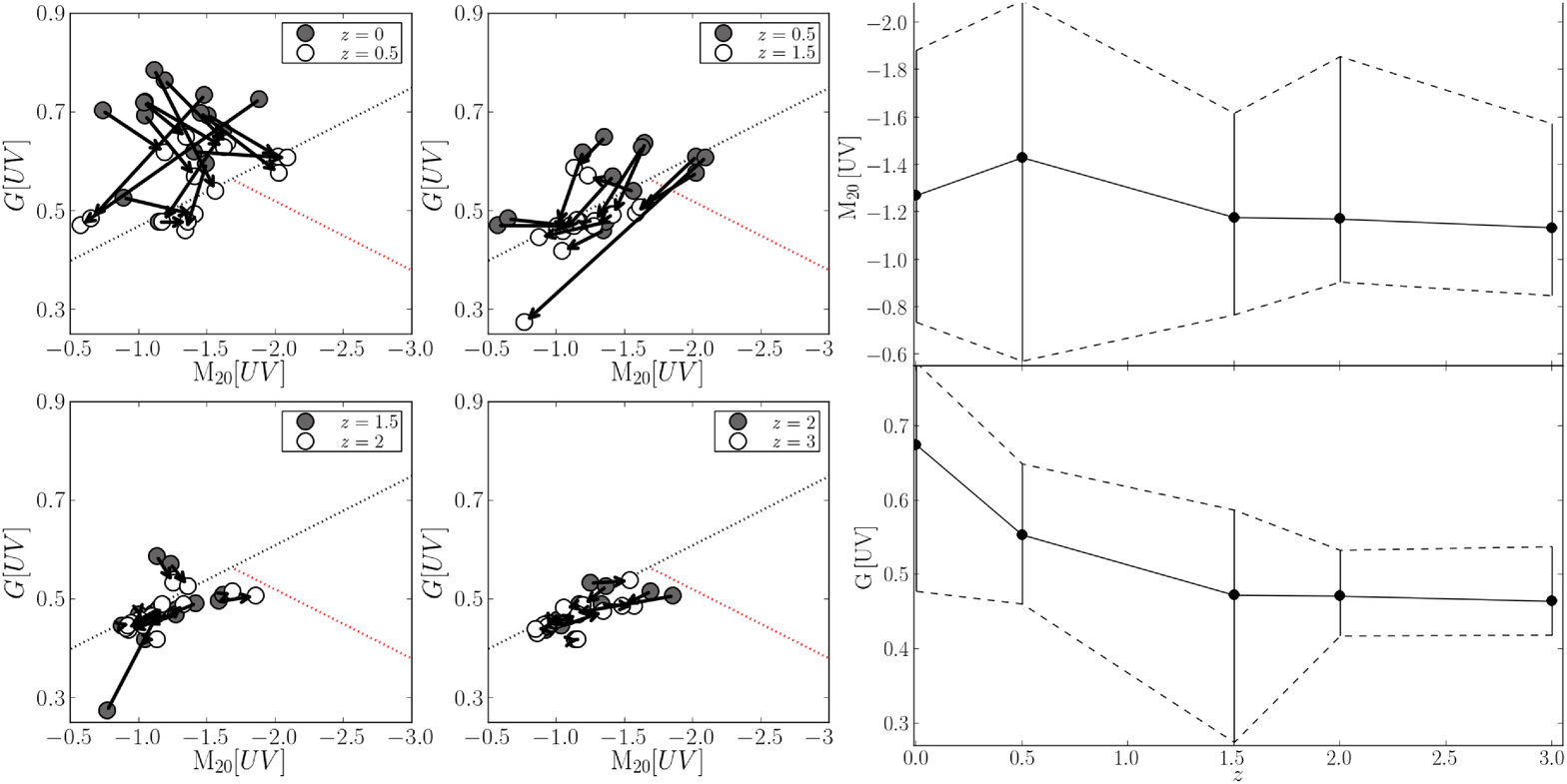}
\caption{
The four panels on the left show the G-\M20~ values for the FUV rest-frame {\it redshifted} images. For each redshift bin, we connect the G-\M20 values from an object's lower redshift (grey) to its higher redshift (open circles) as defined in the legend. The two panels on the right show the average G and \M20 values with simulated redshift (black circles). The vertical lines span the minimum and maximum values for that redshift bin. The dashed and solid lines illustrate the trend for G to decrease and flatten with increasing redshift, while \M20 remains relatively stable.
}\label{fig7}
\end{figure*}

}

\def\figh{
\begin{figure*}
\plotone{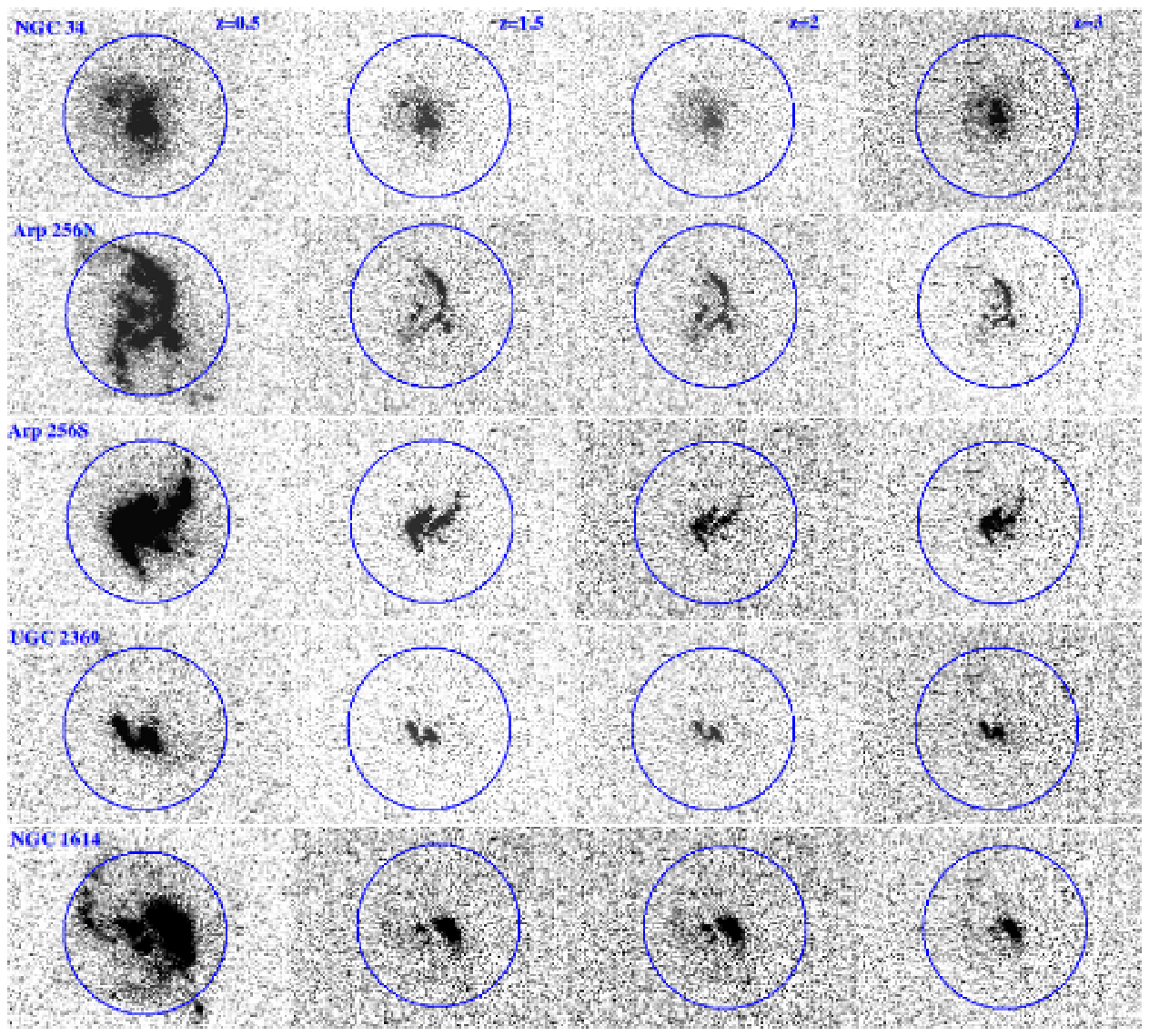}
\caption{
Five of 15 objects artificially redshifted in the rest-frame FUV from $0.5\leq z \leq 3$ as labeled, and simulated to the depth of the UDF in the bands from left to right: F220W ($z=$0.5), B ($z=$1.5,2), and V ($z=$3). All circles are $1.5\arcsec$ in radius.
}\label{fig7a}
\end{figure*}

}
\def\figi{
\begin{figure*}
\plotone{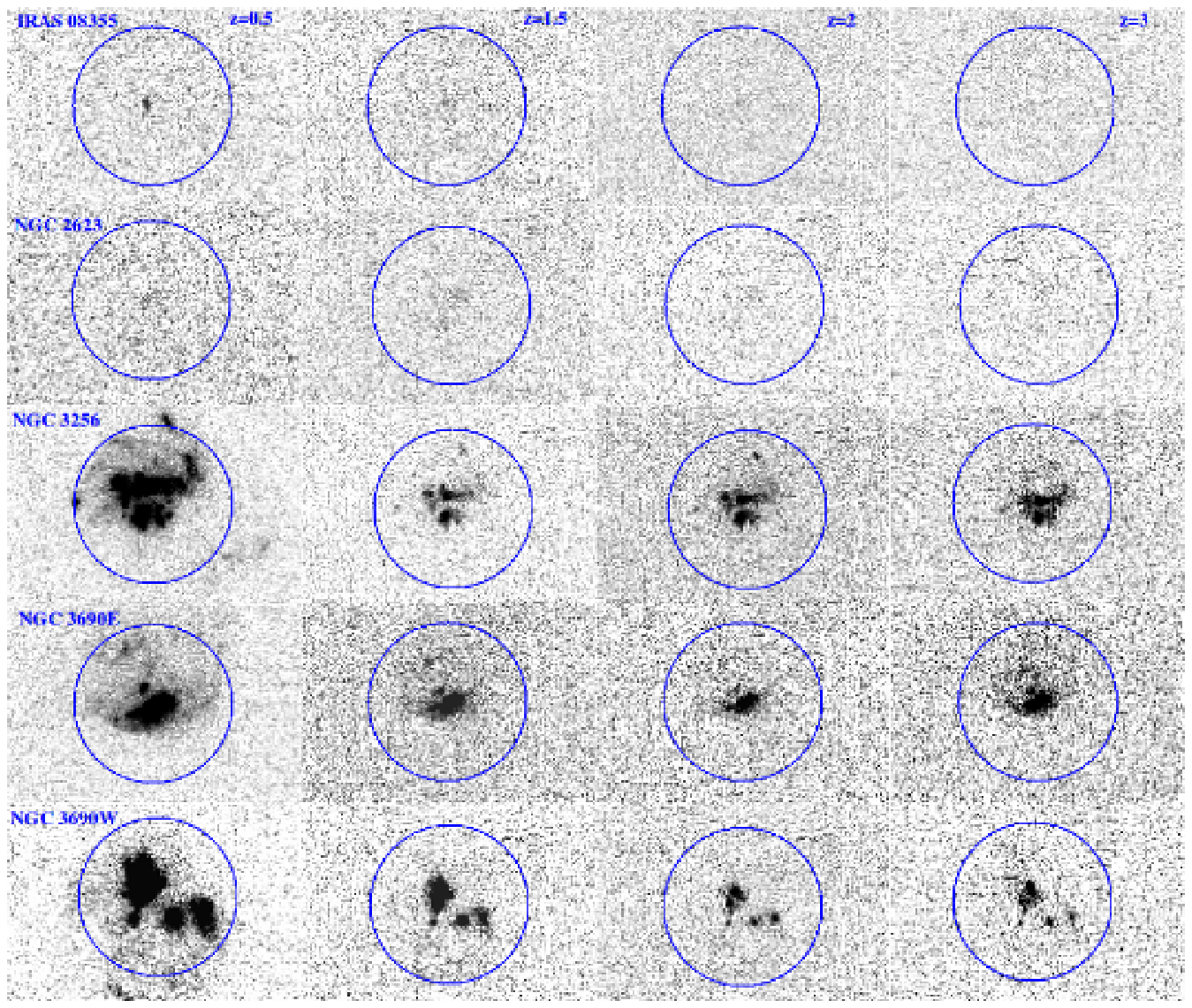}
\caption{Continued from Figure \ref{fig7a}
}\label{fig7b}
\end{figure*}

}

\def\figj{
\begin{figure*}
\plotone{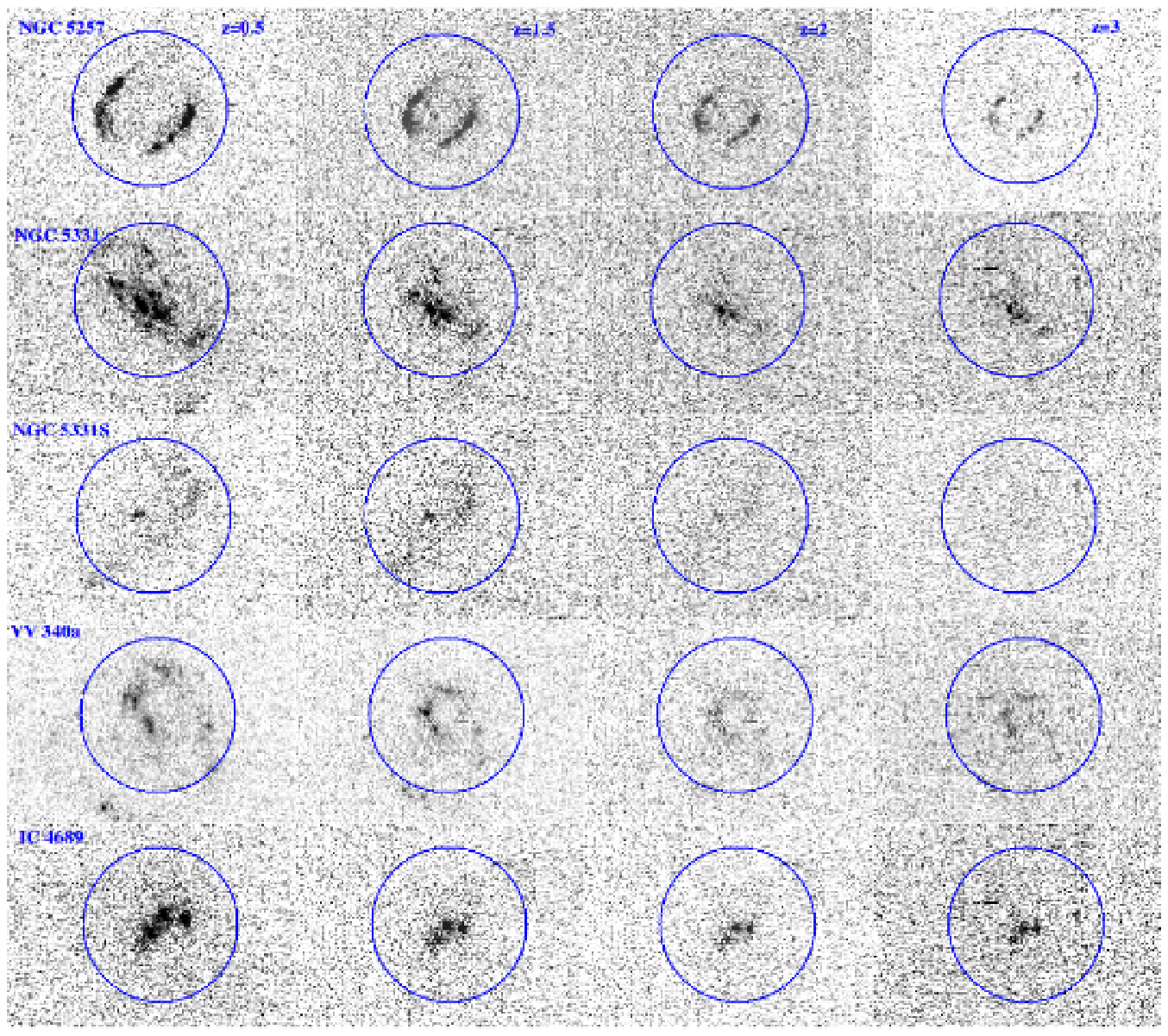}
\caption{Continued from Figure \ref{fig7a}
}\label{fig7c}
\end{figure*}

}

\def\figk{
\begin{figure*}
\plotone{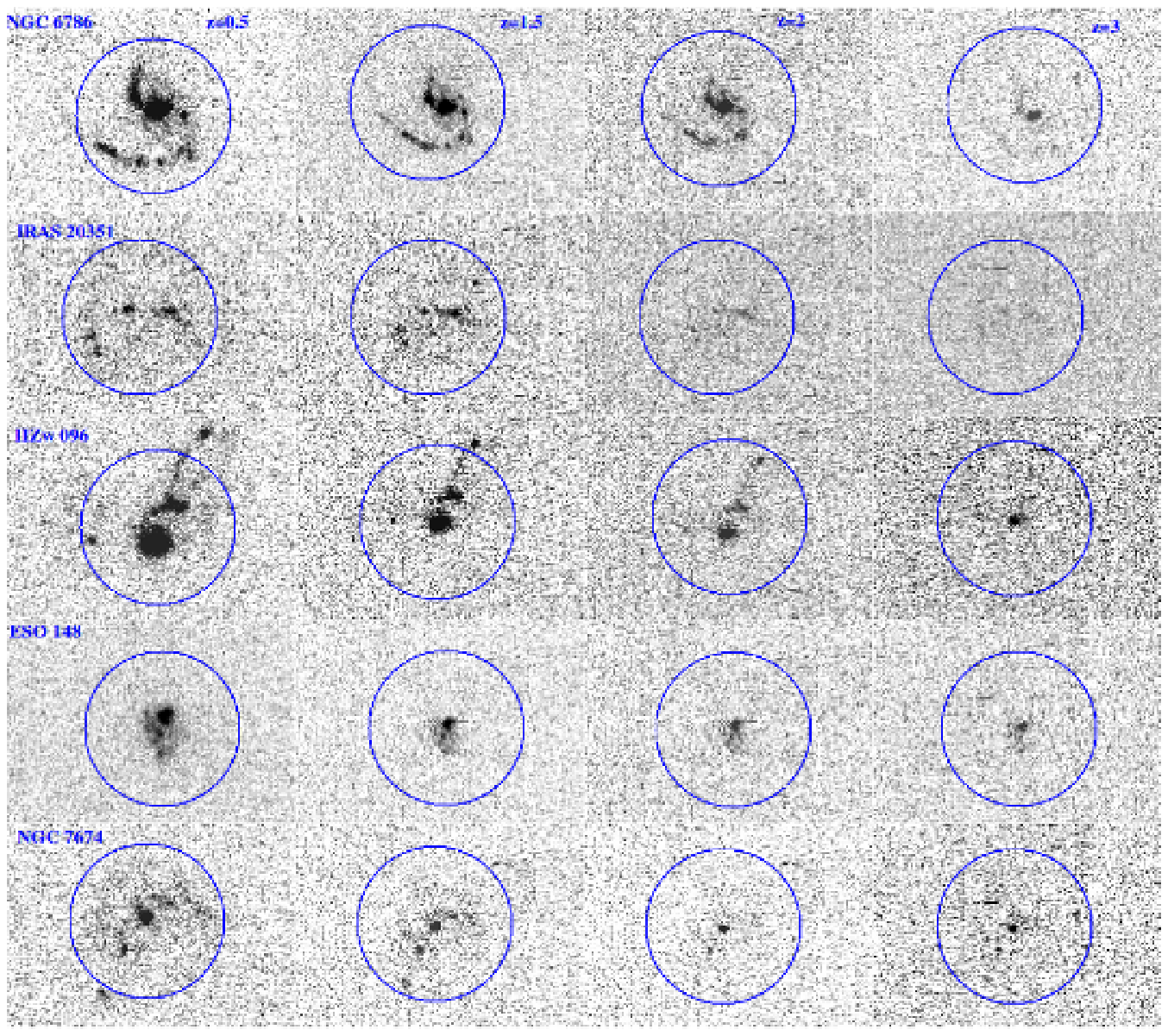}
\caption{Continued from Figure \ref{fig7a}
}\label{fig7d}
\end{figure*}

}
\def\tablea{

\begin{deluxetable*}{lcccccc}

\tabletypesize{\scriptsize}
\tablecaption{General Properties of 20 LIRGs \label{tableA}}

\tablehead{
\colhead{} & 
\colhead{RA} & 
\colhead{DEC} & 
\colhead{D\tablenotemark{a}} &
\colhead{s\tablenotemark{a}} &
\colhead{$\log\, \mathrm{L_{IR}/L_\odot}$\tablenotemark{b}} &
\colhead{Merger Stage\tablenotemark{c}} \\ 
\colhead{} & 
\colhead{} & 
\colhead{} & 
\colhead{(Mpc)} &
\colhead{(pc pix$^{-1}$)} &
\colhead{} & 
\colhead{} 
} 

\startdata
NGC 34            		& 00h11m06.6s & -12d06m26s & 82.8 &30   &  11.43 &5\\
Arp 256N    & 00h18m50.1s & -10d21m42s &41  & 39 &  10.45 &2 \\
Arp 256S          		& 00h18m50.9s & -10d22m37s & 114.4&41   &  11.44 &2\\
UGC 02369         		& 02h54m01.8s & +14d58m14s & 131.7&47  & 11.66&2\\
NGC 1614          		& 04h33m59.9s & -08d34m44s & 67.3 &24  & 11.65& 5\\
IRAS 08355-4944& 08h37m01.8s & -49d54m30s & 39.5 &39   &  11.62 & 3\\
NGC 2623          		& 08h38m24.1s & +25d45m17s & 44.0 &28   &  11.60 & 5\\
NGC 3256          		& 10h27m51.3s & -43d54m14s & 43.2 &14   &  11.64 & 5\\
NGC 3690E    & 11h28m33.6s & +58d33m47s & 95.7 &16   &  11.58 &3\\
NGC 3690W         		& 11h28m31.0s & +58d33m41s & 105.6&15  &  11.67 &3\\
NGC 5257          		& 13h39m52.9s & +00d50m24s & 152.4&34  &  11.31 &2\\
NGC 5331          		& 13h52m16.3s & +02d06m11s & 188.3&50   &  11.02 &2\\
NGC 5331S         		& 13h52m16.2s & +02d06m03s & 122.1&49  &  11.54 &2\\
VV 340a            		& 14h57m00.8s & +24d37m04s & 78.2 &50   &  11.66 &1\\
IC 4689           		& 18h13m40.3s & -57d44m53s & 109.4&25 &  10.88 &2\\
NGC 6786        		& 19h10m53.9s & +73d24m37s & 142.2&38  & 11.23 &1\\
IRAS 20351+2521   		& 20h37m17.7s & +25d31m38s & 69.7 &50   & 11.61 &0\\
IIZW  96          		& 20h57m23.9s & +17d07m39s & 142.3&54   &  11.77 &2 \\
ESO 148-IG 002    		& 23h15m46.7s & -59d03m16s & 139.5&63.6 &  12.06 &4 \\
NGC 7674      & 23h27m56.7s & +08d46m45s & 138.5&44  &  11.54 &1 \\

\enddata

\tablenotetext{a}{Distances (D) were obtained from the
NASA/IPAC Extragalactic Database (NED). The pixel scale in parsec per pixel, s, is based on the NICMOS H-band pixel scale ($0.076 \rm\, arcsec\, pix^{-1}$) and the distance, D.}
\tablenotetext{b}{$\rm L_{IR}$ calculations are described in \citet{dia13}.}
\tablenotetext{c}{I-band merger-stages classified by \citet{haa11}: 0 = single galaxy, galaxy with minor companion; 1 = separate galaxies, disks symmetric (intact), no tails; 
2 = progenitor galaxies distinguishable, disks asymmetric or amorphous, tidal tails; 3 = two nuclei in common envelope; 4 = double nuclei with tidal tail; 5 =  single or obscured nucleus, long prominent tails;
6 = single or obscured nucleus, disturbed central morphology, short faint tails or tails absent, shells.}
\end{deluxetable*}
}

\def\M20{$M_{20}$}

\def\lir{$\rm L_{IR}$}

\def\M20{$M_{20}$}

\begin{document}

\title{The FUV to Near-IR Morphologies of Luminous Infrared Galaxies in the GOALS Sample}

\author{Petty, S.M.\altaffilmark{1}, Armus, L.\altaffilmark{2}, Charmandaris, V.\altaffilmark{3,4,5}, Evans, A.S.\altaffilmark{6,7}, Le Floc'h, E.\altaffilmark{8}, Bridge, C.\altaffilmark{9}, D\'iaz-Santos, T.\altaffilmark{2}, Howell, J.H.\altaffilmark{2}, Inami, H.\altaffilmark{10}, Psychogyios, A.\altaffilmark{3}, Stierwalt, S.\altaffilmark{6}, Surace, J.A.\altaffilmark{2} }

  \altaffiltext{1}{Dept. of Physics, Virginia Tech Blacksburg, VA, 24061}
  \altaffiltext{2}{Spitzer Science Center, California Institute of Technology, Pasadena, CA 91125, USA}
  \altaffiltext{3}{Department of Physics, University of Crete, GR-71003, Heraklion, Greece}
\altaffiltext{4}{Institute for Astronomy, Astrophysics, Space Applications \& Remote Sensing,
National Observatory of Athens, GR-15236, Penteli, Greece}
\altaffiltext{5}{Chercheur Associ\'e, Observatoire de Paris, F-75014,  Paris, France}
     \altaffiltext{6}{Department of Astronomy, University of Virginia, Charlottesville, VA 22904}
     \altaffiltext{7}{National Radio Astronomy Observatory, Charlottesville, VA 22903}
   \altaffiltext{8}{CEA-Saclay, Orme des Merisiers, Bat. 709, 91191 Gif-sur-Yvette, France}
  \altaffiltext{9}{Div. of Physics, Math \& Astronomy, California  Institute of Technology, Pasadena, CA 91125}
  \altaffiltext{10}{National Optical Astronomy Observatory, Tucson, AZ 85719}

\begin{abstract}
 We compare the morphologies of a sample of 20 LIRGs from the Great Observatories All-sky LIRG Survey (GOALS) in the FUV, B, I and H bands, using the Gini (G) and \M20 parameters to quantitatively estimate the distribution and concentration of flux as a function of wavelength. HST images provide an average spatial resolution of $\sim 80$ pc. While our LIRGs can be reliably classified as mergers across the entire range of wavelengths studied here, there is a clear shift toward more negative \M20 (more bulge-dominated) and a less significant decrease in G values at longer wavelengths. We find no correlation between the derived  FUV  G-\M20 parameters and the global measures of the IR to FUV flux ratio, IRX.  Given the fine resolution in our HST data, this suggests either that the UV morphology and IRX are correlated on very small scales, or that the regions emitting the bulk of the IR emission emit almost no FUV light. We use our multi-wavelength data to simulate how merging LIRGs would appear from $z\sim0.5-3$ in deep optical and near-infrared images such as the HUDF, and use these simulations to measure the G-\M20 at these redshifts. Our simulations indicate a noticeable decrease in G, which flattens at $z\geq 2$ by as much as 40\%, resulting in mis-classifying our LIRGs as disk-like, even in the rest-frame FUV.  The higher redshift values of \M20 for the GOALS sources do not appear to change more than about 10\% from the values at $z\sim0$. The change in G-\M20 is caused by the surface brightness dimming of extended tidal features and asymmetries, and also the decreased spatial resolution which reduced the number of individual clumps identified. This effect, seen as early as $z\sim 0.5$, could easily lead to an underestimate of the number of merging galaxies at high-redshift in the rest-frame FUV. 
 
\end{abstract}

\section{Introduction}
Luminous Infrared Galaxies (LIRGs) are defined as a class of infrared-dominant galaxies with $\rm L_{IR}\, (8-1000\mu m) \geq 10^{11} \, L_\odot$ in \citet{hou85} based on observations by the Infrared Astronomical Satellite (IRAS). The most luminous members of this class, Ultra-LIRGs (ULIRGs $\rm L_{IR} = 10^{12}-10^{13} \, L_\odot$) and Hyper-LIRGs (HyLIRGs $\rm L_{IR} > 10^{13} \, L_\odot$), have been the central subclass in much of the literature on LIRGs, where the focus has been on their merger rates and contribution to cosmic infrared background from $z\gtrsim 1$ \citep[e.g.,][]{san96,vei02,got05,lon06}. The lower luminosity local LIRGs ($\rm L_{IR} = 10^{11}-10^{12} \, L_\odot$) have more diverse morphologies, compared to ULIRGs, which tend to be late-stage major mergers \citep[see][and references therein]{san96}. The diverse LIRG morphology, coupled with the lower fraction of major mergers relative to ULIRGs, may reflect the broader evolutionary pathways of a galaxy to become a LIRG at low redshift. While most LIRGs are starburst-dominated \citep{pet11,sti13}, many contain AGN. This, combined with the wide range of merger morphologies, makes them excellent laboratories for studying the AGN-SB connection under a variety of physical conditions.

Rough correlations between luminosity and merger stage have been extensively discussed \citep[e.g.,][]{vei02, got05, bri07, kar10b, pet11, haa11},  and are generally attributed to episodes of enhanced star formation during the close passage and final merger of two gas-rich galaxies.  Most low redshift LIRGs are found in systems which have undergone strong tidal perturbations due to the merging of gas-rich disk galaxies \citep[][]{arm87, san88a, san88b, mur96, mur01, far01, bri07, kar10b}. At higher luminosities, the number density of ULIRGs increases with redshift \citep[e.g.,][]{sch76,san96,kim98,lon06}. Approximately 50\% of the cosmic infrared background is due to LIRGs which dominate star formation rates at $z>1$ \citep[][]{lef05,cap07,mag09,ber11,mag13}. \citet{kar12} conclude that 24\% of main sequence \citep{elb11} and 73\% of starburst ULIRGs are merging or interacting at $z\sim2$. Over half of their sample have AGN, while $\sim25\%$ of these AGN are classified as mergers or interacting, and mostly lie on the main sequence. 

Mid-IR diagnostics of the entire Great Observatories All-sky Lirg Survey \citep[GOALS;][]{arm09} sample indicate that AGN are present in 18\% of LIRGs, and 10\% contribute more than 50\% of the total IR luminosity \citep[][]{pet11}. \citet{sti13} find that the starburst dominated LIRGs resemble $z\sim2$ starburst dominated submillimeter galaxies in their mid-IR spectral properties. While the spectroscopic features accurately separate the contributions of star formation versus AGN to the total \lir, the morphological distinction between a dust-shrouded AGN versus dusty starburst galaxies is not definitive. 
 
Even though LIRGs are powered by luminous starbursts and often AGN the UV properties of LIRGs have received far less scrutiny, since much of this radiation is absorbed by dust. Because of their extreme luminosities, the small fraction of the UV radiation that escapes can make LIRGs strong sources of UV flux \citep{how10}. When observed in the UV-optical wavelengths, dust may create the appearance of a very disturbed or post-merger morphology for an underlying disk structure. The location of the dust and gas, with respect to the young stars, can have a large impact on the appearance of a LIRG in the UV, the amount of radiation that escapes into the intergalactic medium, and the overall global properties. It is natural then to assume that the wide range in integrated IR to UV ratios and UV slopes seen in LIRGs might be related to the small-scale UV morphologies.  

Previous morphology surveys in the rest-frame UV have primarily focused on optically bright starbursts and disks to gain an unobscured view of the distribution of star-forming regions \citep[e.g.,][]{con03,lotz06,petty09,law07a,law12}. The galaxies generally appear as disk-like and/or made up of compact irregular structures. Visually, there is a substantial difference between the UV and near-IR morphologies in dusty galaxies. The brightest clusters or even the nuclei in the IR can be extremely faint in the UV, making classifications highly uncertain \citep[e.g., ][]{hib97,mir98,char04}. Quantitative morphological studies of local LIRGs covering the full range of observed morphologies from the rest-frame UV through the near-IR are few in number. Evidence from studies of high-redshift, star-forming galaxies suggest that there is no quantifiable change in the basic morphological parameters across these wavelengths \citep{law12}. However, the issue of wavelength dependence morphologies is not resolved in the literature. Also, most of the galaxies are selected in the UV or optical, and they may show very different properties to those galaxies selected in the far-infrared.

In this paper, we use a large sample of local LIRGs observed in the rest-frame far-UV through the near-infrared with HST to: 1) measure the UV to H-band morphologies in a consistent quantitative manner, using Gini and \M20; 2) use the infrared excess (IRX) to explore correlations of the UV morphology with global IR and UV properties; 3) demonstrate how redshift and surface brightness dimming can quantitatively affect the apparent UV morphology of IR-selected merging galaxies at $z \sim 3$. Here we present and analyze the ACS FUV images of a small sample of 20 LIRGs, compare these to the visual and near-IR morphologies of the same galaxies on the same scales, and relate quantitative morphological measures to the global IR and FUV properties. This is the first smaller local LIRG subsample from which complete FUV to far-IR HST data are available. Our sample of 20 LIRGs is described in Section \ref{sample}. In Section \ref{qmethods} we discuss our methodology. In Section \ref{results}, we begin by treating each galaxy individually, and then as a set determine how G-\M20 values are affected by dust in multiple wavelengths. Where applicable, we adopt a $\Lambda$CDM cosmology, ($\rm \Omega_m, \, \Omega_\Lambda, \, H$) = (0.27, 0.73, 71).

\section{Sample Selection }\label{sample}

The  Great Observatories All-sky LIRG Survey \citep[GOALS;][]{arm09} is a multi-wavelength imaging and spectroscopic study of 202 of the most luminous infrared-selected galaxies in the local Universe, including observations from X-ray to far-IR. It is an ideal sample for comparing the effects of dust on the visual and quantitative morphologies from the UV to near-IR with the superb HST spatial resolutions. The GOALS targets are drawn from the IRAS Revised Bright Galaxy Sample \citep[RBGS; ][]{san03}, a complete sample of 629 galaxies with IRAS 60 $\mu$m flux densities S$_{60} > 5.24$ Jy, covering the full sky above Galactic latitudes $|b| > 5$ degrees. The 629 IRAS RBGS galaxies have a median redshift of $z = 0.008$ with a maximum of $z = 0.088$ and include 180 LIRGs and 22 ULIRGs that define the GOALS sample. 

As part of a pilot program to obtain FUV images of the GOALS galaxies, we obtained ACS/F140LP images of a GOALS galaxy sample with $\rm L_{IR} > 10^{11.4} L_{\odot}$. A secondary goal was to study the ages and distributions of young star clusters in the sample (Evans, in prep). Therefore, the sources for UV imaging were selected from the existing ACS B, and I-band data to have significant numbers of bright ($\rm 23 < B < 21$ AB) star clusters. This resulted in a sample of 22 LIRGs, five of which are part of pairs, so 27 individual galaxies were observed in the FUV with the  Advanced Camera for Surveys (ACS) High Resolution Channel. Only 20 of these have corresponding NICMOS imaging, which is the number of individual galaxies in our final sample.

We have complete UV to near-IR HST imaging for these 20 LIRGs. The UV and optical ACS filters include the SBC F140LP (UV), and F435W (B) and F814W (I). The near-IR images were observed with the Near-Infrared Camera and Multi-Object Spectrometer (NICMOS) F160W filter (H). The HST FUV (F140LP) data were obtained with the ACS SBC over the period 2007 July 21 to 2009 August 2 (PID 11196; PI: A.S. Evans). The SBC has a $34\arcsec \times 31\arcsec$ field of view (FOV), with a pixel scale of 0.032 arcsec/pixel. The data were taken in ACCUM mode using the PARALLELOGRAM dither pattern. We further reduced the SBC data using standard STScI routines: the images were calibrated using CALACS, then distortion-corrected and combined using PyDrizzle. The PyDrizzle routine also removes any cosmic rays and bad pixels present in the images. We refer the reader to \citet{haa11} for the NICMOS data and data reduction description. The B and I-band data and data reduction are described in \citet{kim13}.
\tablea
In Table \ref{tableA}, we list the sample and several derived properties (distance, pixel scale, \lir, and merger stage) for each of the galaxies in our sample. The IR luminosities have a range of $10.45<\rm \log \, L_{IR}/L_{\odot}<12.06$. The two IR luminosities below $10^{11}\ \rm L_\odot$ are for individual components of interacting systems. The sample does represent a diverse visual morphology at different stages of merging from the 202 LIRGs in the GOALS catalog: Column 8 lists the I-band visually classified merger stages from \citet{haa11}. The merger stages are described in Table \ref{tableA}. Only one LIRG is classified for type 0 and 4 stages (i.e., IRAS 20351: single nucleus with a minor companion, and ESO 148: double nuclei with a tidal tail). Eight LIRGs are classified as type 2, having distinguishable individual galaxies with disks and tidal tails. The LIRGs span distances of 40-188 Mpc with a mean of 106 Mpc. Given the HST NICMOS resolution ($0.\arcsec16$), this provides an average spatial resolution of 80 pc \citep[also see ][ for a description of the NICMOS PSF]{haa11}.

\figaa
\figab
\figac
\figad

\section{Morphological diagnostics} \label{qmethods}

The use of Gini (G) and \M20 for quantifying the morphology of a a galaxy is described in detail in \citet{lotz04} and \citet{lotz06}. The Gini coefficient was originally developed as an economic indicator for determining the distribution of money in a society \citep{gini12}. Applied to Astrophysics, G is correlated with the concentration of light of a source as well as with it surface brightness, and is defined as 
\begin{equation}
\rm{G} = \frac{1}{\displaystyle |\bar{f}|n(n-1)}\sum_{i=1}^{n}(2i-n-1)|f_{i}|.
\end{equation}
where $|\bar{f}|$, and $|f_i|$ are the absolute average flux, and $i$th pixel flux from n total pixels, respectively. Gini may typically have values between 0.3-0.8, where larger values are indicative of a more centralized, concentrated light distribution.  \M20 is a logarithmic ratio giving the normalized second-order moment of the brightest pixels whose integrated flux adds up to 20\% of the total flux of the galaxy and is defined by
 \begin{equation}
M_{20} \equiv {log10}\left(\frac{\sum_i M_i}{M_{\rm tot}}\right).
\end{equation}
$M_i$ is given by
\begin{equation}
M_i = f_i [(x_i-x_c)^2 + (y_i-y_c)^2] \, {\rm with} \, \sum_i f_i > 0.2 f_{\rm tot},
\end{equation}
where $f_i$ is the flux at position $x_i$ and $y_i$ of the $i$th pixel; $x_c$ and $y_c$ is the galaxy center image coordinates. The pixels are rank ordered by flux for calculating both parameters. \M20 is more negative for a galaxy with 20\% of the light concentrated in a bulge or single nucleus. We use the G-\M20 parameter space to identify morphological types. \citet{lotz08b} defines G-\M20 values to classify mergers, early type and disk like morphologies:
\begin{equation}\label{eq:merger}
{\rm Mergers: G >-0.14} M_{20} + 0.33 \rm ;
\end{equation}
\begin{equation}\label{eq:eso}
{\rm E/S0/Sa: G \leq -0.14} M_{20} +0.33{\rm; \, and \, G >0.14} M_{20} +0.80\rm; 
\end{equation}
\begin{equation}\label{eq:sb}
{\rm Sb-Ir: G \leq -0.14} M_{20} +0.33 {\rm; and G \leq} 0.14 M_{20} +0.80.
\end{equation}
Although these classifications are based on UV-optical morphologies, we use these same regions to define  UV-, B-, I- and H-band morphologies, for comparison. We show in Section \ref{results} that these region definitions are relatively robust at all wavelengths for all but a few cases.

The standard procedure is to calculate the total flux within 1.5 times the Petrosian radius ($r_p$) for G-\M20 \citep[see][]{lotz04}. Many clumpy, star-forming galaxies are elongated, due to their irregular shape, so choosing an ellipsoid can more accurately measure the distribution of light in the defined pixel area. While this minimizes the inclusion of background flux, it works for images with enough background to more accurately determine the surface brightness. For comparing across the wavelengths, the size of the radius is limited to the smallest field of view, i.e., the NICMOS FOV, and in most cases using a radius of 1.5$r_p$  to calculate G-\M20 extends beyond this. Thus, we elected to use circular apertures of $7.8-15.4 \arcsec$ (within the NICMOS FOV) for all of the images, including those at shorter wavelengths. We show the NICMOS H-band footprint in Figures \ref{fig1}-\ref{fig1c}.

We subtracted an average sky flux from all frames before calculating G-\M20. To measure the background, we used the IRAF \textit{imexamine} tool\footnote{Image Reduction and Analysis Facility: http://iraf.net}. Random errors were estimated by changing the H-band center (center of mass, $x_c,\ y_c$) by 5 arcseconds in random directions about the actual galaxy centroid, measuring the G and \M20 using the same radius, and calculating the standard error. This method accounts for errors in defining the center between wavelengths that would change \M20.  An additional error was added in quadrature, to account for systematics, since the derived values of Gini and \M20 depend on the precise background subtraction, binning and psf convolution to common sampling and resolution across wavelengths. This is particularly true when measuring these parameters using deep, wide-field images of complex objects whose structures can change by a great deal as one moves from the UV through the NIR.  While some of these dependencies are captured within the random uncertainties that we have propagated through the analysis of the data, we have added, in quadrature, an average systematic uncertainty of 0.045 in Gini and 0.1 in \M20 to the measured parameters for each object. These were derived from repeatedly varying the background subtraction, binning and radii of the measurements.

The images in the FUV-, B- and I-bands were re-binned (degraded) to match the H-band image, due to the H-band having the largest pixel scale and PSF (point spread function). We deconvolved each image with the band-specific PSF and convolved with the H-band PSF. This allows for the morphologies to be directly comparable, since the pixel scales and PSFs can lead to substantial changes in morphologies. We obtain spatial resolutions on average of $\sim60-100$ pc based on the 0.076 arcsec pixel$^{-1}$ NICMOS pixel angular resolution.

\section{The Multi-wavelength Morphologies of 20 LIRGs: Results}\label{results} 

In this section, we focus on G- and \M20-based classification of our local LIRG sample from the UV through near-IR. We compare the derived morphologies in the FUV-, B-, I-and H-bands to assess any changes in G and \M20, we compare with visual classifications, and the IR to FUV flux ratio.

Visually, LIRGs change in morphology from the FUV to H bands (i.e., from clumpy major mergers to smooth disks with well-defined centers), and the purpose of this analysis is to determine how, at HST resolutions, we can discriminate between these classifications with our prescribed morphological diagnostic. Our sample is wholly comprised of interacting systems at various stages. Table \ref{tableA} lists the stages based on NIR \citep{haa11}. Since LIRGs are dusty, we expect their morphologies to vary with wavelength, and therefore their position in the G-M20 parameter space. However, the precise manner of how these changes affect G-\M20 is not clear. Exploring this morphology parameter space of local LIRGs, will be useful for comparisons with other galaxy samples, as the majority of automated morphology studies focus on higher redshift ($z>0.2$) galaxies in the rest-frame UV and optical.

\subsection{Probing FUV-, B-, I- and H-band LIRG Morphologies with G-\emph{M$_{20}$}}\label{resultsa}
Many of the LIRGs in the FUV show resolved and unresolved clumps, as well as extended features, characterized by a complex morphology similar to the optical and near-IR. Figures \ref{fig1}-\ref{fig1c} show cutouts of the FUV-, B-, I- and H-band images (left to right) of the entire sample of 20 galaxies. The box regions outline the H-band footprint, which in all cases is slightly smaller than the FOV in the UV and optical ACS data. Individual G-\M20 plots are included to the right of the individual galaxy cutouts. Each point in the G-\M20 plane is derived from one of the FUV, B, I and H images (FUV- and H-band points are annotated as UV and H, respectively). We also show the boundaries between the regions where mergers, E-Sa, and Sb-Ir galaxies typicall fall \citep[G-\M20 boundary, hereafter; ][]{lotz08b}. 

Based on the G-\M20 boundaries, the multi-wavelength morphologies are consistently in the merger region for 13/20 objects in all four bands. We do not consider those on the merger boundary to have changed class, because of uncertainties inherent in the derivation of G-\M20 for a particular source (see Section \ref{qmethods} and Figure \ref{fig3}). For example, IC 4689 shows an FUV-band  G-\M20 value on the merger/non-merger boundary, but we do not consider it to have changed type from a merger to Sb-Ir, or E-Sa type. 

\figc

Seven LIRGs reside in different regions in G-\M20 space as a function of wavelength: UGC 02369, IRAS 08355, NGC 3256, NGC 3690E, VV 340a, NGC 6786 and NGC 7674. These objects have visual classifications of type 2, 3, 3, and 1, respectively. 

UGC 02369 has a very irregular shape with a compact nucleus and an adjacent extended bar-like structure bright in the FUV- and B-bands, and fades to look like a spiral arm in the I- and H-bands. The values for G-\M20 are in the merger boundary in the FUV, hover along the merger/non-merger border for B- and I-bands, and shift to the border between the E-S0/Sb-Ir region in the H-band. 

In the FUV, IRAS 08355 is entirely dominated by the bright nucleus. As a result, the FUV image of this galaxy places it squarely inside the Sb-Ir boundary in the G-\M20 plane. The light for both nuclei in the post-merger envelope is revealed redward of FUV. The B, I, and H values for G-\M20 do vary, staying just within the merger boundary. 

NGC 3256 shows striking differences in symmetry between wavelengths. the FUV- and B-band images show a significant portion of the lower disk is  masked by dust. The I-band reveales a disk-like pattern with many clumpy star-forming regions. The H-band shows a bright nucleus and an asymmetric distribution of light along the arms. G-\M20 values are quite different for each wavelength, but stay within the merger boundary in the FUV-, B- and  H-bands, while the I-band values are within the Sb-Ir boundary. Given the errors, the I-band value is arguably on the border. 

In NGC 3690E the G-\M20 for FUV and H are consistent within the merger boundary. The B and I values are distinctly different and are located within the Sb-Ir boundary. Visually the FUV-, B-, and I-band images are very similar, and in this case the G-\M20 values do not reflect that. The disk-like nature is clear in the H-band.

In all wavelengths VV 340a shows a disk-like structure. In the FUV, the nucleus is shrouded by dust and the morphology takes on a pattern of clumpy, tightly-wound spiral arms. In the longer wavelengths, the nucleus is revealed and the morphology resembles more of a normal disk. This behavior is reflected in the G-\M20 values where the points shift from the merger boundary in the FUV-band, to the Sb-Ir boundary in the B-, I- and H-bands.  

The loosely bound spiral arms in NGC 6786 are visible in each wavelength. However, the FUV- and B-bands do show a more clumpy, disjointed pattern from the nucleus to the extended arms. In the I- and H-bands the emission appears more smooth and symmetric with a bright bulge and bar. All but the H-band G-\M20 are in the merger boundary.

NGC 7674 appears very disk-like in all bands. FUV and B reveal clumpy disk arms, where much of the light is veiled by structured dust lanes. The G-\M20 points are in the merger boundary for the FUV, and the Sb-Ir boundary for all other bands, where the nucleus is much brighter and the spiral arms more smoothly distributed.

Four of these cases (UGC 02369, VV 340a, NGC 6786, and NGC 7674),  the movement in the G-\M20 plane with wavelength is as expected, namely a straightforward progression from merger to a more regular morphology at longer wavelengths. The other three objects have a more complicated progression. An interesting result is that majority of our sample do not follow the expected pattern of merger to more regular morphology with increased wavelength, given the copious amounts of dust in LIRGs.

Figure \ref{fig3} shows the G-\M20 values for the FUV-, B-, I- and H-bands in four panels. The dashed lines indicate the G-\M20 boundaries for different morphological types. We included errors based on the methods described in Section \ref{qmethods}. The averages for Gini are $\rm <G> = 0.64\pm0.02, \ 0.62\pm0.02, \ 0.60\pm0.02, \ and \ 0.67\pm0.02$; and the averages for \M20 are $\rm <\mathit{M}_{20}>\rm = -1.24\pm0.19, \ -1.33\pm0.17, \ -1.53\pm0.20, \ and \ -1.89\pm0.27$ in the FUV-, B-, I- and H-bands, respectively. The averages clearly show that \M20 becomes more negative with longer wavelengths. Gini narrowly hovers around $0.6$ with a standard deviation of $\approx 0.1$. While the classification does not change on average with wavelength, the shift toward a more negative \M20 from FUV to the H-band is consistent with a bright bulge being revealed in the near-IR. 

Five of the 20 galaxies are known hosts to AGN: NGC 34, NGC 3690E, NGC 3690W, NGC 6786, and NGC 7674. They are labeled in Figure \ref{fig3} as yellow pentagons. A bright nuclear point source can affect the location of an object in the G-\M20 plane, but these five AGN do not appear to significantly affect the location of the host galaxies in Figure \ref{fig3}. It is expected that the definition of the center of flux highly biases the quantitative morphology; therefore, galaxies with highly obscured nuclei would exhibit the biggest differences between wavelength. 
\fige
Since most of the LIRGs are mergers, the morphological classification, due to both stellar tidal features and dust, should change as the merger progresses.  In Figure \ref{fig5} we explore the variation of G and \M20 in our sample as a function of merger stage \citep[see Table \ref{tableA} and ][]{haa11}. The black diamonds (FUV), black triangles (H), grey squares (B) and grey circles (I) are the mean G and \M20 values of all galaxies in a given merger type. The averages for all of the values, regardless of wavelength are show by the solid black line. Note that there is only one type 0, IRAS 20351, and one type 4, ESO 148, so these values are for the individual galaxy. Type 0 are single galaxies with a minor companion, and type 4 are double nuclei with a tidal tail. The overall trend (black line) indicates that Gini, and \M20 don't vary significantly as a function of merger stage. There is a slight rise in Gini in the later stages of mergers, but the uncertainties are large and the effect is subtle.

One way to quantify the affects of dust on the emitted UV radiation is by referring to the ratio of infrared to FUV flux (infrared excess, IRX) and UV slope \citep[$\beta$;][]{how10}. In general, LIRGs extend over a large range in IRX-$\beta$ as shown in Figure \ref{fig3a}.  Since IRX is driven by the fraction of bolometric emission originating from regions which become thick in the UV, due to the presence of high quantities of dust, we might expect a correlation between IRX, $\beta$ and UV morphology. 
 
Sixteen of the objects in our sample with available UV data are a subset of the LIRGs found in \citet{how10}, who analyzed the IRX-$\beta$ relation for the entire GOALS sample. Although the current sample does not span the full range of IRX and $\beta$ seen among local LIRGs, the subset of 16 galaxies studied here does span nearly two orders of magnitude in IRX (0.3-2.3), and a significant range in $\beta$ (-1 to +0.5). 
\figca
\figcb
Since IRX measures the global ratio between the FIR and FUV light, and this light originates from star-forming complexes in the nuclei and disk of each LIRG, we might expect a correlation between IRX and the morphologies of the galaxies, particularly in the FUV as measured with G and \M20. Figure \ref{fig4} shows IRX versus \M20 (top row) and G (bottom row) for each band, UV-H. The grey shaded areas are values attributed to a more disturbed light distribution; disturbed galaxies generally have G $>0.57$ and \M20 $>-1.7$, which is the intersection of the G-\M20 boundary lines (see Figure \ref{fig4}). Aside from the general trend for our sample to have more negative \M20 values in the H-band, there is no obvious correlation between IRX and Gini or \M20 among LIRGs. Galaxies with extremely large IRX do not have \M20 values significantly different than the rest of the sample. While there may be a tendency for high IRX systems to have lower Gini, the UV panel shows a larger scatter in Gini among those galaxies with large IRX, even though all sources with G$\lesssim0.5$ also have IRX $>2.0$.

\M20~ has a larger scatter into the more bulge-dominated region in the I and H-bands (\M20~ is more negative). G does not change between wavelengths. The Gini outlier in B and I is NGC 3690E, and in H is Arp 256N. Here we find no correlation between IRX and G-\M20 classification in the FUV, or other wavelengths. 

\subsection{Simulations of high-$z$ LIRGs in the rest-frame FUV }\label{redshift}

Numerous papers analyze the apparent morphology of distant galaxies and how they are affected by resolution and depth \citep[e.g.,][]{hib97,lotz04}.  Similarly, the often clumpy nature of UV imaging data can make even normal, star-forming galaxies appear disturbed and irregular. Our sample of rest-frame HST FUV data of local mergers enables us to simulate the appearance of high-z mergers in the optical bands. In particular, using our sample, we simulate how the LIRGs in the SBC FUV rest-frame would appear in ultra-deep field (UDF) images at $z=$ 0.5, 1.5, 2, and 3, in order to examine if local LIRGs would be identified as galaxy mergers at these redshifts. 
\figh
\figi
\figj
\figk

\figg

The simulated images are generated by applying the same method described in \citet{gia96a} \citep[see][for more recent applications]{petty09,bri10a}. Briefly, we rebin and convolve the SBC images according to the pixel scale and PSF of the bands at the depth of the UDF. We use the F220W ($z=0.5$), B ($z=1.5$, 2), and V-band (F606W; $z=3$) filter responses and detector characteristics (0.03$\arcsec \, \rm pix^{-1}$), corresponding with the redshifted FUV rest-frame band. The limiting AB magnitudes are 25.9, 28.7, and 29 for the F220W, B and V bands, respectively \citep{gia04,bec06}. We rebinned the $z\sim0$ FUV images to the calculated angular size scales at each redshift, convolved with the instrument PSFs \citep[produced by Tiny Tim;][]{kri11}, and added background from each band in the UDF to  match the noise levels \citep{gia04,bec06}. We exclude five LIRGs from this analysis that were not detected after being redshifted to $z = 3$: NGC 5331S, VV 340a, NGC 2623, IRAS 20351+2521, IRAS 08355-4944, leaving 15 for the full range of redshifts to analyze. 

The radii used for measuring G and M20 for the high-redshift simulations have the same projected physical size ($6-10.8$ kpc) used for the $z\sim0$ values. This allows for a direct comparison between the nearby and simulated higher redshift values of G and \M20. To check the sensitivity of the high-$z$ G-\M20 measurements, we adjusted the radius for each galaxy at the incremental radii: $r/2, 2r$. At r/2, Gini decreases on average by $\sim 0.02$, while \M20 increases (becomes less negative) on average by $\sim0.2$; at 2r Gini increases on average by $\sim 0.01$, and \M20 decreases by $\sim0.1$. The differences are well within the uncertainty ranges estimate described in Section \ref{qmethods}.

In Figures \ref{fig7a}-\ref{fig7d}, the simulated UDF rest-frame FUV images of the LIRGs are shown at each redshift. The circles have radii of 1.5$\arcsec$ ($\rm \sim9-13$ kpc in the redshift range considered here), and are for comparison purposes. The majority of the LIRGs retain the brightest clump structures, while low surface-brightness areas are lost to noise, or some of the clumps blend in the rebinning process. The galaxies that appear most affected by this --NGC 6786, IIZw 96, ESO 148, NGC 7674-- have been mostly reduced to a single bright clump that would be regarded as the entire galaxy. Structural details are lost for $z\gtrsim 2$, and they might be mistaken for compact UV-bright galaxies. This is in agreement with the results of \citet{hib97} performed on a smaller sample.

The morphologies as estimated by the G-\M20 values further confirm this result. The left four panels in Figure \ref{fig7} compares G-\M20 between each redshift bin, starting with the original values at $z \sim 0$.  We also include the G-\M20 boundary lines defining the expected G-\M20 morphology types as in Figure \ref{fig3}. By the first step in the redshift sequence at $z=0.5$, the majority of objects have moved to the merger boundary. As the redshift increases, the galaxies transition almost entirely into the Sb-Ir region, and out of the merger area.

We also show \M20 and G versus redshift in the two right panels. While \M20 may vary for the individual galaxies at each redshift as shown in the G-\M20 plots, the overall ranges of \M20 show no trend with redshift. However, the G values decrease significantly, until $z\gtrsim1.5$ where the values flatten to G$\sim0.45$. Several factors, including the reduction of spatial resolution from 80 pc to 1.36 kpc, in addition to dimming of tails and other extended features, have a systematic affect on both \M20 and G. The KS-tests between the original, or $z\sim0$ and redshifted G values are inconsistent with them arising from the same sample (P values $\ll 0.1\%$). For \M20 the KS-tests give P values $>10\%$ for $z=$0.5, 1.5 and 3 when compared with the original LIRG \M20 value. 

Overall, at higher redshifts, most merging LIRGs in the FUV-rest-frame would appear as compact Sb-Ir galaxies. Visually, some may appear clumpy and disturbed (Figures \ref{fig7a}-\ref{fig7}), but many of the extended features are absent, which is one key attribute to a merger classification. Based on our small, yet diverse sample, merging galaxies at $z \gtrsim 2$ can appear to be similar to normal disk galaxies when observed at visual wavelengths.
 
 NGC 5257 presents an interesting case, which points to the opposite possibility; namely, the classification of disk galaxies with irregular star formation as mergers. NGC 5257 is a member of an interacting pair \citep{haa11}, where the value for G-\M20 in the NIR sit on the border between irregular and normal disk galaxies. The G-M20 values as measured in the optical and UV however are well within the irregular/merger region. When redshifted to $z>1$, NGC 5257 follows the same trend toward smaller G as the other LIRGs in the sample, but straddles the dividing line between the merger and disk galaxy regions in the G-M20 plane. Therefore, although this source is a member of an interacting pair, the G-\M20 classification suggests a distorted disk, driven by the asymmetric star formation evident in our images. In this case the classification is correct, but is entirely driven by the unusual star formation in the disk of NGC 5257 itself. If galaxies with similar asymmetries are common at higher redshifts, they would also appear as mergers even if the star formation is driven by other processes, such as gas accretion and disk instabilities. 

Overall, the largest effect on G-\M20 is that G decreases significantly, due to the decrease in signal to noise, $(1+z)^4$ surface brightness dimming, and decrease in spatial resolution from pixel rebinning, during the redshifting process. Even at UDF depths, the surface brightness dimming removes extended tidal features and other asymmetries. The pixel rebinning significantly reduces the number of resolved individual clumps, concentrating the brightest pixels.

\section{Discussion and Conclusions}\label{discussion}

We have presented a study of the G-\M20 derived morphologies in local sample of LIRGS, observed with Hubble from the the FUV through near-IR at an average spatial resolutions of 80 pc. Visually, the distribution of individual star-forming clusters in the LIRGs appear to depend on wavelength. Whether the morphology of a dusty galaxy may be effectively identified based on UV-optical observations has been explored previously. \citet{mir98} analyzed the double nuclei in the Antennae galaxies with HST optical and mid-IR (ISO). They found that the brightest mid-IR cluster is dark in the optical wavelengths, and used these results to postulate that this would bias high redshift observations. \citet{hib97} studied four nearby irregular galaxies using images in the rest-frame FUV, B- and V-bands and simulated them to look like Hubble Deep Field observations at higher redshifts ($0.5<z<2.5$). They demonstrated that it was not possible to recover the global properties of the galaxies after $z>1.5$. They further argued that the rest-frame UV was not sensitive to an older, more general stellar population, such as the shorter, optical wavelengths. 

Our results indicate that quantitative G-\M20 measurements do not effectively reflect this wavelength dependence. While the location of mergers in the G-\M20 plane move significantly from the FUV through the visual and near-IR, they do not typically cross the boundaries that define merging systems to those occupied by regular galaxies (Spiral or Elliptical). Merging LIRGs stay in the same G-\M20 classification independent of wavelength. This result may be a function of both resolution and depth, but it is nonetheless an observed property of these data. For our dusty merger sample, the G-\M20 classifications seem robust from the FUV through the NIR. 

However, there are some obvious changes where G and \M20 behave differently with wavelength: \M20 becomes more negative and G increases (within a very narrow range), which moves the G-\M20 values into the upper right corner of the merger region with increasing wavelength. These results are consistent with a dusty central bulge being revealed in the near-IR and dominating the brightest pixels. Specifically, this effect is driving the G-\M20 changes for NGC 34 and NGC7674, which visually become more bulge-/disk-like. Some of our galaxies harbor AGN, which do not seem to affect their G-\M20 values compared to the starbursts in our sample. Even though \citet{gab09} show that the AGN can affect the host morphology, we do not find this to be the case for nearby LIRGs.

Both G and \M20 are very good at selecting merger/LIRGs regardless of rest-frame wavelength at $z\sim0$. However, Gini and \M20 do not correlate clearly with merger stage, although there is a slight rise in Gini with late stage mergers in our sample. This is promising for selecting merger candidates from large catalogs, where the rest-frame does not need to be taken into account within a reasonable confidence level. 

We also compare G-\M20 with IRX and find no correlation. This lack of correlation between the IRX, G, and \M20 parameters suggests that a significant change in the amount of dust reprocessing of the FUV emission (over two orders in magnitude in IRX) is not changing the overall UV morphology in a systematic way across the sample, as revealed in G-\M20.  This in turn suggests that the dust is not preferentially affecting the brightest UV knots, but rather that it is either (1) evenly distributed across the galaxies, or (2) is surrounding and obscuring the bulk of the FUV emission from a small number of regions which in turn emit most of the FIR light.  Since we know that the starbursts and the dust emission in LIRGs becomes more concentrated as the luminosity increases in the later merger stages, this result is consistent with the warm dust which drives LIRGs to higher IRX being concentrated around the nucleus. An increase in IRX would seem to provide a smaller contribution of the brightest clumps to the total UV flux. We are probing the \ion{H}{2} region scales, implying that any correlation between IRX and morphology would be occurring on smaller scales.

Due to the levels of dust obscuration causing low-surface brightness, rest-frame UV and optical surveys of higher redshift infrared luminous galaxies are more rare \citep[e.g.,][]{con03,lotz06,rav06,petty09,con11,law12,hol12}. \citet{elb07} visually classify the morphologies of LIRGs at $z\sim1$, and find that only a third have signatures of being in a major merger. In comparison, \citet{kar12} find that $\simeq 24\%$ of ULIRGs at $z\sim2$ are mergers.

Based on our simulation of LIRGs at increasing redshift, we detect a consistent loss of information due to surface brightness dimming, removing the tidal tails and other faint asymmetric features in late-stage merging systems. Plus, a loss of resolution of individual clumps occurs, because of the pixel rebinning. The results from redshifting the LIRGs up to $z=3$ suggest that G is highly sensitive to these effects. However, \M20 appears to be relatively stable to the redshift simulations with an average difference of $\sim4\%$ from the range of values it takes at $z\sim0$. Moreover, our analysis shows that G decreases as much as 40\% over the range of values. It is likely that the loss of resolution (clumps combine with pixel rebinning), and surface brightness dimming of the faint, extended features dominate the change, skewing G-\M20 toward a smoother disk-like parameter space. In other words, merging galaxies will look more disk-like at $z\gtrsim2 $ in the G-\M20 plane, even in the rest-frame FUV.

While our results show that mergers, when seen at high redshift, can often appear more ``regular'' in the visual and UV than they really are, the converse is also true: galaxies can appear to be mergers when in fact they are irregular, as in the case of NGC 5257.

A number of authors have already discussed the significantly lower fraction of mergers seen among high-$z$ galaxies.  At fixed luminosities this is certainly a factor, since galaxies will lie along a rising main sequence. However, our data suggest that the exact fraction of mergers at $z \gtrsim 0.5$ can also be underestimated. In \citet{wan12}, the authors measure G-\M20~on a color-selected sample of extremely red objects (mostly disks) at $z\sim2$ in the rest-frame optical. They find a clean separation in G-\M20 space between the quiescent and dSFGs, and infer that 30\% of star-forming galaxies and 70\% of quiescent galaxies are disks, claiming that this argues against a significant role for mergers in building galactic mass. However, we show that merging systems at high redshift with faint extended features may be classified as disks. \citet{petty09} also find that nearby merger-like galaxies simulated at $z\sim1.5$ and 4, in the rest-frame UV, have more disk-like G-\M20 values. 

There are some discrepancies in determining the morphology-wavelength dependence, or k-correction, particularly for estimating mergers. Understanding this factor is important for determining merger-fractions at higher redshifts. The results we find for this LIRG sample are similar to the basic conclusion of \citet{ove10}, who compare nearby Lyman-break analogs (LBAs), LBGs at $z\sim3$-4, and BzK galaxies at $z\sim2$ in the rest-frame UV and optical at HUDF depths. They show that the extended features of the LBAs disappear after degrading the images to the BzK and LBG sensitivities (simulating higher redshifts), implying that the rest-frame UV is particularly susceptible to the removal of extended features.

This effect could have an impact on studies of mergers as the major source of mass building which rely purely  on morphological classifications of high-$z$ galaxies. In particular, observations and simulations show that disks can survive and reform after a major wet merging event \citep[e.g., ][]{rob06,rob08,hop09,gov09}. Even though our results do not suggest that all star-forming $z\gtrsim 0.5$ disks or recent starbursts are powered by wet, major mergers, it is clear that the number of major mergers may be underestimated when using G-\M20, and other quantitative measures to recognize merging systems at higher redshifts in the rest-frame UV.  In particular, although our sample is small, we find that $\sim50\%$ would appear disk-like if observed at UDF-like depths with HST by $z\sim0.5$, and nearly all of them are classified as disk-like by $z\sim2$. At low-redshift, galaxies look more irregular in the UV due to dust and the dominance of hot, young stars, but when observed at high-redshift this trend can be reversed, and even mergers can look disk-like. In future papers we will explore these trends with a larger sample of local mergers observed from the UV through NIR with HST, and analyze the star-forming clumps as a function of merger stage and other global parameters, such as dust temperature and SED shape.

\acknowledgments Support for Program number HST-GO-11196 was provided by NASA through a grant from the Space Telescope Science Institute, which is operated by the Association of Universities for Research in Astronomy, Incorporated, under NASA contract NAS5-26555. VC would like to acknowledge partial support from the EU FP7 Grant PIRSES-GA-2012-316788. This research has made use of the NASA/IPAC Extragalactic Database (NED) which is operated by the Jet Propulsion Laboratory, California Institute of Technology, under contract with the National Aeronautics and Space Administration.

\bibliographystyle{apj}

\begin{thebibliography}{58}
\expandafter\ifx\csname natexlab\endcsname\relax\def\natexlab{1}\fi

\bibitem[{{Armus} {et~al.}(1987){Armus}, {Heckman}, \& {Miley}}]{arm87}
{Armus}, L., {Heckman}, T., \& {Miley}, G. 1987, \aj, 94, 831

\bibitem[{{Armus} {et~al.}(2009){Armus}, {Mazzarella}, {Evans}, {Surace},
  {Sanders}, {Iwasawa}, {Frayer}, {Howell}, {Chan}, {Petric}, {Vavilkin},
  {Kim}, {Haan}, {Inami}, {Murphy}, {Appleton}, {Barnes}, {Bothun}, {Bridge},
  {Charmandaris}, {Jensen}, {Kewley}, {Lord}, {Madore}, {Marshall},
  {Melbourne}, {Rich}, {Satyapal}, {Schulz}, {Spoon}, {Sturm}, {U}, {Veilleux},
  \& {Xu}}]{arm09}
{Armus}, L., {Mazzarella}, J.~M., {Evans}, A.~S., {et~al.} 2009, \pasp, 121,
  559

\bibitem[{{Beckwith} {et~al.}(2006){Beckwith}, {Stiavelli}, {Koekemoer},
  {Caldwell}, {Ferguson}, {Hook}, {Lucas}, {Bergeron}, {Corbin}, {Jogee},
  {Panagia}, {Robberto}, {Royle}, {Somerville}, \& {Sosey}}]{bec06}
{Beckwith}, S.~V.~W., {Stiavelli}, M., {Koekemoer}, A.~M., {et~al.} 2006, \aj,
  132, 1729

\bibitem[{{Berta} {et~al.}(2011){Berta}, {Magnelli}, {Nordon}, {Lutz}, {Wuyts},
  {Altieri}, {Andreani}, {Aussel}, {Casta{\~n}eda}, {Cepa}, {Cimatti}, {Daddi},
  {Elbaz}, {F{\"o}rster Schreiber}, {Genzel}, {Le Floc'h}, {Maiolino},
  {P{\'e}rez-Fournon}, {Poglitsch}, {Popesso}, {Pozzi}, {Riguccini},
  {Rodighiero}, {Sanchez-Portal}, {Sturm}, {Tacconi}, \& {Valtchanov}}]{ber11}
{Berta}, S., {Magnelli}, B., {Nordon}, R., {et~al.} 2011, \aap, 532, A49

\bibitem[{{Bridge} {et~al.}(2010){Bridge}, {Carlberg}, \& {Sullivan}}]{bri10a}
{Bridge}, C.~R., {Carlberg}, R.~G., \& {Sullivan}, M. 2010, \apj, 709, 1067

\bibitem[{{Bridge} {et~al.}(2007){Bridge}, {Appleton}, {Conselice}, {Choi},
  {Armus}, {Fadda}, {Laine}, {Marleau}, {Carlberg}, {Helou}, \& {Yan}}]{bri07}
{Bridge}, C.~R., {Appleton}, P.~N., {Conselice}, C.~J., {et~al.} 2007, \apj,
  659, 931

\bibitem[{{Caputi} {et~al.}(2007){Caputi}, {Lagache}, {Yan}, {Dole},
  {Bavouzet}, {Le Floc'h}, {Choi}, {Helou}, \& {Reddy}}]{cap07}
{Caputi}, K.~I., {Lagache}, G., {Yan}, L., {et~al.} 2007, \apj, 660, 97

\bibitem[{{Charmandaris} {et~al.}(2004){Charmandaris}, {Le Floc'h}, \&
  {Mirabel}}]{char04}
{Charmandaris}, V., {Le Floc'h}, E., \& {Mirabel}, I.~F. 2004, \apjl, 600, L15

\bibitem[{{Conselice}(2003)}]{con03}
{Conselice}, C.~J. 2003, \apjs, 147, 1

\bibitem[{{Conselice} {et~al.}(2011){Conselice}, {Bluck}, {Ravindranath},
  {Mortlock}, {Koekemoer}, {Buitrago}, {Gr{\"u}tzbauch}, \& {Penny}}]{con11}
{Conselice}, C.~J., {Bluck}, A.~F.~L., {Ravindranath}, S., {et~al.} 2011,
  \mnras, 417, 2770

\bibitem[{{D{\'{\i}}az-Santos} {et~al.}(2013){D{\'{\i}}az-Santos}, {Armus},
  {Charmandaris}, {Stierwalt}, {Murphy}, {Haan}, {Inami}, {Malhotra},
  {Meijerink}, {Stacey}, {Petric}, {Evans}, {Veilleux}, {van der Werf}, {Lord},
  {Lu}, {Howell}, {Appleton}, {Mazzarella}, {Surace}, {Xu}, {Schulz},
  {Sanders}, {Bridge}, {Chan}, {Frayer}, {Iwasawa}, {Melbourne}, \&
  {Sturm}}]{dia13}
{D{\'{\i}}az-Santos}, T., {Armus}, L., {Charmandaris}, V., {et~al.} 2013, \apj,
  774, 68

\bibitem[{{Elbaz} {et~al.}(2007){Elbaz}, {Daddi}, {Le Borgne}, {Dickinson},
  {Alexander}, {Chary}, {Starck}, {Brandt}, {Kitzbichler}, {MacDonald},
  {Nonino}, {Popesso}, {Stern}, \& {Vanzella}}]{elb07}
{Elbaz}, D., {Daddi}, E., {Le Borgne}, D., {et~al.} 2007, \aap, 468, 33

\bibitem[{{Elbaz} {et~al.}(2011){Elbaz}, {Dickinson}, {Hwang},
  {D{\'{\i}}az-Santos}, {Magdis}, {Magnelli}, {Le Borgne}, {Galliano},
  {Pannella}, {Chanial}, {Armus}, {Charmandaris}, {Daddi}, {Aussel}, {Popesso},
  {Kartaltepe}, {Altieri}, {Valtchanov}, {Coia}, {Dannerbauer}, {Dasyra},
  {Leiton}, {Mazzarella}, {Alexander}, {Buat}, {Burgarella}, {Chary}, {Gilli},
  {Ivison}, {Juneau}, {Le Floc'h}, {Lutz}, {Morrison}, {Mullaney}, {Murphy},
  {Pope}, {Scott}, {Brodwin}, {Calzetti}, {Cesarsky}, {Charlot}, {Dole},
  {Eisenhardt}, {Ferguson}, {F{\"o}rster Schreiber}, {Frayer}, {Giavalisco},
  {Huynh}, {Koekemoer}, {Papovich}, {Reddy}, {Surace}, {Teplitz}, {Yun}, \&
  {Wilson}}]{elb11}
{Elbaz}, D., {Dickinson}, M., {Hwang}, H.~S., {et~al.} 2011, \aap, 533, A119

\bibitem[{{Farrah} {et~al.}(2001){Farrah}, {Rowan-Robinson}, {Oliver},
  {Serjeant}, {Borne}, {Lawrence}, {Lucas}, {Bushouse}, \& {Colina}}]{far01}
{Farrah}, D., {Rowan-Robinson}, M., {Oliver}, S., {et~al.} 2001, \mnras, 326,
  1333

\bibitem[{{Gabor} {et~al.}(2009){Gabor}, {Impey}, {Jahnke}, {Simmons}, {Trump},
  {Koekemoer}, {Brusa}, {Cappelluti}, {Schinnerer}, {Smol{\v c}i{\'c}},
  {Salvato}, {Rhodes}, {Mobasher}, {Capak}, {Massey}, {Leauthaud}, \&
  {Scoville}}]{gab09}
{Gabor}, J.~M., {Impey}, C.~D., {Jahnke}, K., {et~al.} 2009, \apj, 691, 705

\bibitem[{{Giavalisco} {et~al.}(1996){Giavalisco}, {Livio}, {Bohlin},
  {Macchetto}, \& {Stecher}}]{gia96a}
{Giavalisco}, M., {Livio}, M., {Bohlin}, R.~C., {Macchetto}, F.~D., \&
  {Stecher}, T.~P. 1996, \aj, 112, 369

\bibitem[{{Giavalisco} {et~al.}(2004){Giavalisco}, {Ferguson}, {Koekemoer},
  {Dickinson}, {Alexander}, {Bauer}, {Bergeron}, {Biagetti}, {Brandt},
  {Casertano}, {Cesarsky}, {Chatzichristou}, {Conselice}, {Cristiani}, {Da
  Costa}, {Dahlen}, {de Mello}, {Eisenhardt}, {Erben}, {Fall}, {Fassnacht},
  {Fosbury}, {Fruchter}, {Gardner}, {Grogin}, {Hook}, {Hornschemeier}, {Idzi},
  {Jogee}, {Kretchmer}, {Laidler}, {Lee}, {Livio}, {Lucas}, {Madau},
  {Mobasher}, {Moustakas}, {Nonino}, {Padovani}, {Papovich}, {Park},
  {Ravindranath}, {Renzini}, {Richardson}, {Riess}, {Rosati}, {Schirmer},
  {Schreier}, {Somerville}, {Spinrad}, {Stern}, {Stiavelli}, {Strolger},
  {Urry}, {Vandame}, {Williams}, \& {Wolf}}]{gia04}
{Giavalisco}, M., {Ferguson}, H.~C., {Koekemoer}, A.~M., {et~al.} 2004, \apjl,
  600, L93

\bibitem[{Gini(1912)}]{gini12}
Gini, C. 1912, Memorie di metodologia statistica

\bibitem[{{Goto}(2005)}]{got05}
{Goto}, T. 2005, \mnras, 360, 322

\bibitem[{{Governato} {et~al.}(2009){Governato}, {Brook}, {Brooks}, {Mayer},
  {Willman}, {Jonsson}, {Stilp}, {Pope}, {Christensen}, {Wadsley}, \&
  {Quinn}}]{gov09}
{Governato}, F., {Brook}, C.~B., {Brooks}, A.~M., {et~al.} 2009, \mnras, 398,
  312

\bibitem[{{Haan} {et~al.}(2011){Haan}, {Surace}, {Armus}, {Evans}, {Howell},
  {Mazzarella}, {Kim}, {Vavilkin}, {Inami}, {Sanders}, {Petric}, {Bridge},
  {Melbourne}, {Charmandaris}, {Diaz-Santos}, {Murphy}, {U}, {Stierwalt}, \&
  {Marshall}}]{haa11}
{Haan}, S., {Surace}, J.~A., {Armus}, L., {et~al.} 2011, \aj, 141, 100

\bibitem[{{Hibbard} \& {Vacca}(1997)}]{hib97}
{Hibbard}, J.~E., \& {Vacca}, W.~D. 1997, \aj, 114, 1741

\bibitem[{{Holwerda} {et~al.}(2012){Holwerda}, {Pirzkal}, \& {Heiner}}]{hol12}
{Holwerda}, B.~W., {Pirzkal}, N., \& {Heiner}, J.~S. 2012, \mnras, 427, 3159

\bibitem[{{Hopkins} {et~al.}(2009){Hopkins}, {Cox}, {Younger}, \&
  {Hernquist}}]{hop09}
{Hopkins}, P.~F., {Cox}, T.~J., {Younger}, J.~D., \& {Hernquist}, L. 2009,
  \apj, 691, 1168

\bibitem[{{Houck} {et~al.}(1985){Houck}, {Schneider}, {Danielson},
  {Neugebauer}, {Soifer}, {Beichman}, \& {Lonsdale}}]{hou85}
{Houck}, J.~R., {Schneider}, D.~P., {Danielson}, G.~E., {et~al.} 1985, \apjl,
  290, L5

\bibitem[{{Howell} {et~al.}(2010){Howell}, {Armus}, {Mazzarella}, {Evans},
  {Surace}, {Sanders}, {Petric}, {Appleton}, {Bothun}, {Bridge}, {Chan},
  {Charmandaris}, {Frayer}, {Haan}, {Inami}, {Kim}, {Lord}, {Madore},
  {Melbourne}, {Schulz}, {U}, {Vavilkin}, {Veilleux}, \& {Xu}}]{how10}
{Howell}, J.~H., {Armus}, L., {Mazzarella}, J.~M., {et~al.} 2010, \apj, 715,
  572

\bibitem[{{Kartaltepe} {et~al.}(2010){Kartaltepe}, {Sanders}, {Le Floc'h},
  {Frayer}, {Aussel}, {Arnouts}, {Ilbert}, {Salvato}, {Scoville}, {Surace},
  {Yan}, {Capak}, {Caputi}, {Carollo}, {Cassata}, {Civano}, {Hasinger},
  {Koekemoer}, {Le F{\`e}vre}, {Lilly}, {Liu}, {McCracken}, {Schinnerer},
  {Smol{\v c}i{\'c}}, {Taniguchi}, {Thompson}, {Trump}, {Baldassare}, \&
  {Fiorenza}}]{kar10b}
{Kartaltepe}, J.~S., {Sanders}, D.~B., {Le Floc'h}, E., {et~al.} 2010, \apj,
  721, 98

\bibitem[{{Kartaltepe} {et~al.}(2012){Kartaltepe}, {Dickinson}, {Alexander},
  {Bell}, {Dahlen}, {Elbaz}, {Faber}, {Lotz}, {McIntosh}, {Wiklind}, {Altieri},
  {Aussel}, {Bethermin}, {Bournaud}, {Charmandaris}, {Conselice}, {Cooray},
  {Dannerbauer}, {Dav{\'e}}, {Dunlop}, {Dekel}, {Ferguson}, {Grogin}, {Hwang},
  {Ivison}, {Kocevski}, {Koekemoer}, {Koo}, {Lai}, {Leiton}, {Lucas}, {Lutz},
  {Magdis}, {Magnelli}, {Morrison}, {Mozena}, {Mullaney}, {Newman}, {Pope},
  {Popesso}, {van der Wel}, {Weiner}, \& {Wuyts}}]{kar12}
{Kartaltepe}, J.~S., {Dickinson}, M., {Alexander}, D.~M., {et~al.} 2012, \apj,
  757, 23

\bibitem[{{Kim} \& {Sanders}(1998)}]{kim98}
{Kim}, D., \& {Sanders}, D.~B. 1998, \apjs, 119, 41

\bibitem[{{Kim} {et~al.}(2013){Kim}, {Evans}, {Vavilkin}, {Armus},
  {Mazzarella}, {Sheth}, {Surace}, {Haan}, {Howell}, {D{\'{\i}}az-Santos},
  {Petric}, {Iwasawa}, {Privon}, \& {Sanders}}]{kim13}
{Kim}, D.-C., {Evans}, A.~S., {Vavilkin}, T., {et~al.} 2013, ArXiv e-prints

\bibitem[{{Krist} {et~al.}(2011){Krist}, {Hook}, \& {Stoehr}}]{kri11}
{Krist}, J.~E., {Hook}, R.~N., \& {Stoehr}, F. 2011, in Society of
  Photo-Optical Instrumentation Engineers (SPIE) Conference Series, Vol. 8127

\bibitem[{{Law} {et~al.}(2007){Law}, {Steidel}, {Erb}, {Pettini}, {Reddy},
  {Shapley}, {Adelberger}, \& {Simenc}}]{law07a}
{Law}, D.~R., {Steidel}, C.~C., {Erb}, D.~K., {et~al.} 2007, \apj, 656, 1

\bibitem[{{Law} {et~al.}(2012){Law}, {Steidel}, {Shapley}, {Nagy}, {Reddy}, \&
  {Erb}}]{law12}
{Law}, D.~R., {Steidel}, C.~C., {Shapley}, A.~E., {et~al.} 2012, \apj, 745, 85

\bibitem[{{Le Floc'h} {et~al.}(2005){Le Floc'h}, {Papovich}, {Dole}, {Bell},
  {Lagache}, {Rieke}, {Egami}, {P{\'e}rez-Gonz{\'a}lez}, {Alonso-Herrero},
  {Rieke}, {Blaylock}, {Engelbracht}, {Gordon}, {Hines}, {Misselt}, {Morrison},
  \& {Mould}}]{lef05}
{Le Floc'h}, E., {Papovich}, C., {Dole}, H., {et~al.} 2005, \apj, 632, 169

\bibitem[{{Lonsdale} {et~al.}(2006){Lonsdale}, {Farrah}, \& {Smith}}]{lon06}
{Lonsdale}, C.~J., {Farrah}, D., \& {Smith}, H.~E. 2006, {Ultraluminous
  Infrared Galaxies}, ed. {Mason, J.~W.} (Springer Verlag), 285

\bibitem[{{Lotz} {et~al.}(2006){Lotz}, {Madau}, {Giavalisco}, {Primack}, \&
  {Ferguson}}]{lotz06}
{Lotz}, J.~M., {Madau}, P., {Giavalisco}, M., {Primack}, J., \& {Ferguson},
  H.~C. 2006, \apj, 636, 592

\bibitem[{{Lotz} {et~al.}(2004){Lotz}, {Primack}, \& {Madau}}]{lotz04}
{Lotz}, J.~M., {Primack}, J., \& {Madau}, P. 2004, \aj, 128, 163

\bibitem[{{Lotz} {et~al.}(2008){Lotz}, {Davis}, {Faber}, {Guhathakurta},
  {Gwyn}, {Huang}, {Koo}, {Le Floc'h}, {Lin}, {Newman}, {Noeske}, {Papovich},
  {Willmer}, {Coil}, {Conselice}, {Cooper}, {Hopkins}, {Metevier}, {Primack},
  {Rieke}, \& {Weiner}}]{lotz08b}
{Lotz}, J.~M., {Davis}, M., {Faber}, S.~M., {et~al.} 2008, \apj, 672, 177

\bibitem[{{Magnelli} {et~al.}(2009){Magnelli}, {Elbaz}, {Chary}, {Dickinson},
  {Le Borgne}, {Frayer}, \& {Willmer}}]{mag09}
{Magnelli}, B., {Elbaz}, D., {Chary}, R.~R., {et~al.} 2009, \aap, 496, 57

\bibitem[{{Magnelli} {et~al.}(2013){Magnelli}, {Popesso}, {Berta}, {Pozzi},
  {Elbaz}, {Lutz}, {Dickinson}, {Altieri}, {Andreani}, {Aussel},
  {B{\'e}thermin}, {Bongiovanni}, {Cepa}, {Charmandaris}, {Chary}, {Cimatti},
  {Daddi}, {F{\"o}rster Schreiber}, {Genzel}, {Gruppioni}, {Harwit}, {Hwang},
  {Ivison}, {Magdis}, {Maiolino}, {Murphy}, {Nordon}, {Pannella}, {P{\'e}rez
  Garc{\'{\i}}a}, {Poglitsch}, {Rosario}, {Sanchez-Portal}, {Santini}, {Scott},
  {Sturm}, {Tacconi}, \& {Valtchanov}}]{mag13}
{Magnelli}, B., {Popesso}, P., {Berta}, S., {et~al.} 2013, \aap, 553, A132

\bibitem[{{Meurer} {et~al.}(1999){Meurer}, {Heckman}, \& {Calzetti}}]{meu99}
{Meurer}, G.~R., {Heckman}, T.~M., \& {Calzetti}, D. 1999, \apj, 521, 64

\bibitem[{{Mirabel} {et~al.}(1998){Mirabel}, {Vigroux}, {Charmandaris},
  {Sauvage}, {Gallais}, {Tran}, {Cesarsky}, {Madden}, \& {Duc}}]{mir98}
{Mirabel}, I.~F., {Vigroux}, L., {Charmandaris}, V., {et~al.} 1998, \aap, 333,
  L1

\bibitem[{{Murphy} {et~al.}(1996){Murphy}, {Armus}, {Matthews}, {Soifer},
  {Mazzarella}, {Shupe}, {Strauss}, \& {Neugebauer}}]{mur96}
{Murphy}, Jr., T.~W., {Armus}, L., {Matthews}, K., {et~al.} 1996, \aj, 111,
  1025

\bibitem[{{Murphy} {et~al.}(2001){Murphy}, {Soifer}, {Matthews}, \&
  {Armus}}]{mur01}
{Murphy}, Jr., T.~W., {Soifer}, B.~T., {Matthews}, K., \& {Armus}, L. 2001,
  \apj, 559, 201

\bibitem[{{Overzier} {et~al.}(2010){Overzier}, {Heckman}, {Schiminovich},
  {Basu-Zych}, {Gon{\c c}alves}, {Martin}, \& {Rich}}]{ove10}
{Overzier}, R.~A., {Heckman}, T.~M., {Schiminovich}, D., {et~al.} 2010, \apj,
  710, 979

\bibitem[{{Petric} {et~al.}(2011){Petric}, {Armus}, {Howell}, {Chan},
  {Mazzarella}, {Evans}, {Surace}, {Sanders}, {Appleton}, {Charmandaris},
  {D{\'{\i}}az-Santos}, {Frayer}, {Haan}, {Inami}, {Iwasawa}, {Kim}, {Madore},
  {Marshall}, {Spoon}, {Stierwalt}, {Sturm}, {U}, {Vavilkin}, \&
  {Veilleux}}]{pet11}
{Petric}, A.~O., {Armus}, L., {Howell}, J., {et~al.} 2011, \apj, 730, 28

\bibitem[{{Petty} {et~al.}(2009){Petty}, {de Mello}, {Gallagher}, {Gardner},
  {Lotz}, {Matt Mountain}, \& {Smith}}]{petty09}
{Petty}, S.~M., {de Mello}, D.~F., {Gallagher}, J.~S., {et~al.} 2009, \aj, 138,
  362

\bibitem[{{Ravindranath} {et~al.}(2006){Ravindranath}, {Giavalisco},
  {Ferguson}, {Conselice}, {Katz}, {Weinberg}, {Lotz}, {Dickinson}, {Fall},
  {Mobasher}, \& {Papovich}}]{rav06}
{Ravindranath}, S., {Giavalisco}, M., {Ferguson}, H.~C., {et~al.} 2006, \apj,
  652, 963

\bibitem[{{Robertson} {et~al.}(2006){Robertson}, {Bullock}, {Cox}, {Di Matteo},
  {Hernquist}, {Springel}, \& {Yoshida}}]{rob06}
{Robertson}, B., {Bullock}, J.~S., {Cox}, T.~J., {et~al.} 2006, \apj, 645, 986

\bibitem[{{Robertson} \& {Bullock}(2008)}]{rob08}
{Robertson}, B.~E., \& {Bullock}, J.~S. 2008, \apjl, 685, L27

\bibitem[{{Sanders} {et~al.}(2003){Sanders}, {Mazzarella}, {Kim}, {Surace}, \&
  {Soifer}}]{san03}
{Sanders}, D.~B., {Mazzarella}, J.~M., {Kim}, D., {Surace}, J.~A., \& {Soifer},
  B.~T. 2003, \aj, 126, 1607

\bibitem[{{Sanders} \& {Mirabel}(1996)}]{san96}
{Sanders}, D.~B., \& {Mirabel}, I.~F. 1996, \araa, 34, 749

\bibitem[{{Sanders} {et~al.}(1988{\natexlab{a}}){Sanders}, {Soifer}, {Elias},
  {Madore}, {Matthews}, {Neugebauer}, \& {Scoville}}]{san88a}
{Sanders}, D.~B., {Soifer}, B.~T., {Elias}, J.~H., {et~al.} 1988{\natexlab{a}},
  \apj, 325, 74

\bibitem[{{Sanders} {et~al.}(1988{\natexlab{b}}){Sanders}, {Soifer}, {Elias},
  {Neugebauer}, \& {Matthews}}]{san88b}
{Sanders}, D.~B., {Soifer}, B.~T., {Elias}, J.~H., {Neugebauer}, G., \&
  {Matthews}, K. 1988{\natexlab{b}}, \apjl, 328, L35

\bibitem[{{Schechter}(1976)}]{sch76}
{Schechter}, P. 1976, \apj, 203, 297

\bibitem[{{Stierwalt} {et~al.}(2013){Stierwalt}, {Armus}, {Surace}, {Inami},
  {Petric}, {Diaz-Santos}, {Haan}, {Charmandaris}, {Howell}, {Kim}, {Marshall},
  {Mazzarella}, {Spoon}, {Veilleux}, {Evans}, {Sanders}, {Appleton}, {Bothun},
  {Bridge}, {Chan}, {Frayer}, {Iwasawa}, {Kewley}, {Lord}, {Madore},
  {Melbourne}, {Murphy}, {Rich}, {Schulz}, {Sturm}, {U}, {Vavilkin}, \&
  {Xu}}]{sti13}
{Stierwalt}, S., {Armus}, L., {Surace}, J.~A., {et~al.} 2013, \apjs, 206, 1

\bibitem[{{Veilleux} {et~al.}(2002){Veilleux}, {Kim}, \& {Sanders}}]{vei02}
{Veilleux}, S., {Kim}, D., \& {Sanders}, D.~B. 2002, \apjs, 143, 315

\bibitem[{{Wang} {et~al.}(2012){Wang}, {Huang}, {Faber}, {Fang}, {Wuyts},
  {Fazio}, {Yan}, {Dekel}, {Guo}, {Ferguson}, {Grogin}, {Lotz}, {Weiner},
  {McGrath}, {Kocevski}, {Hathi}, {Lucas}, {Koekemoer}, {Kong}, \&
  {Gu}}]{wan12}
{Wang}, T., {Huang}, J.-S., {Faber}, S.~M., {et~al.} 2012, \apj, 752, 134

\end{thebibliography}
\ajbib

\clearpage

\clearpage

\end{document}